\documentclass[sigplan,nonacm]{acmart}
\AtBeginDocument{}

\usepackage{tikz}
\usepackage{multirow}
\usepackage{makecell}
\usepackage{filecontents}
\usepackage{booktabs}
\usepackage{amsmath}
\usepackage{amsfonts}
\usepackage{pifont} 
\usepackage{epstopdf} 
\usepackage{enumitem}

\usepackage{booktabs}      \usepackage{xcolor}          \usepackage{pifont}          \usepackage{graphicx}        \usepackage{hyphenat}

\usepackage[compact]{titlesec}
\titleformat*{\section}{\large\bfseries}
\titleformat*{\subsection}{\normalsize\bfseries}
\titlespacing*{\section}{0pt}{*1.}{*0.8}
\titlespacing*{\subsection}{0pt}{*0.8}{*0.5}

\DeclareTextFontCommand{\textbfit}{\fontseries\bfdefault \itshape
}

\usepackage{epsfig,makecell}
\usepackage{subfigure}
\usepackage[labelfont=bf,font=normal,skip=1pt]{caption}

\usepackage[ruled, linesnumbered, noend]{algorithm2e}
\usepackage{tikz}

\newcommand{\sys}{\textsc{Apto}}

\newcommand{\eat}[1]{}
\newcommand{\done}[1]{}

\newcommand{\nostitle}[1]{\noindent{\bf #1}}
\newcommand{\stitle}[1]{\vspace{0.8ex}\noindent{\bf #1}}
\newcommand{\etitle}[1]{\vspace{0.5ex}\noindent{\em\underline{#1}}}

\usepackage[normalem]{ulem}
\renewcommand{\headrulewidth}{0pt}
\usepackage[hang,flushmargin]{footmisc}
\usepackage{textcomp}
\usepackage{graphicx}

\DeclareTextFontCommand{\textbfit}{\fontseries\bfdefault
  \itshape
}

\definecolor{mygreen}{rgb}{0,0.6,0}  
\definecolor{mygray}{rgb}{0.5,0.5,0.5}  
\definecolor{mymauve}{rgb}{0.58,0,0.82}

\begin{document}

\title[{\sys}: Self-Adaptive CPU-GPU Hybrid MoE Inference on Shared PCs]{
\LARGE{{\sys}: Self-Adaptive CPU-GPU Hybrid MoE Inference on Shared PCs}}

\author{Zihang Zhong \quad Jiali Wang \quad Rong Chen \vspace{5pt}}
\affiliation{\institution{Institute of Parallel and Distributed Systems, School of Computer Science, Shanghai Jiao Tong University}
  \city{}
  \country{}
}

\makeatletter
\let\ACM@origsection\section
\let\ACM@origsubsection\subsection
\let\ACM@origsubsubsection\subsubsection
\let\ACM@origparagraph\paragraph
\let\ACM@origbaselinestretch\baselinestretch
\makeatother

\date{}
\frenchspacing

\begin{abstract}

CPU/GPU hybrid inference enables Mixture-of-Experts (MoE) models to run 
on memory-constrained PCs by keeping all expert weights in host memory and 
executing activated experts across both processors. 
Existing hybrid systems, however, assume exclusive resource access and 
determine an upfront CPU/GPU partition, 
so CPU contention from colocated applications can severely imbalance 
the two execution paths. 
This paper presents {\sys}, a self-adaptive hybrid inference system that 
applies dynamic scheduling to expert execution. 
The CPU and GPU executors claim experts from opposite ends of a dual-ended queue 
as they progress, allowing the CPU/GPU work split to adapt to their execution progress 
without predicting contention. 
{\sys} further uses affinity-aware scheduling to expose each executor's preferred experts 
at its queue end and cancellable tail racing to mitigate final-expert stragglers. 
Compared with state-of-the-art GPU-only and hybrid baselines, 
{\sys} achieves average prefill and decode speedups of 2.96$\times$ and 1.52$\times$ under standalone execution, 
and 5.97$\times$ and 3.45$\times$ under concurrent CPU workloads. 
It also accelerates real-world coding-agent applications by 2.02$\times$ on average.

\vspace{-1.5mm}
\end{abstract}

\maketitle

\pagestyle{fancy}
\fancyhf{}
\fancyfoot[C]{\thepage}
\renewcommand{\headrulewidth}{0pt}
\renewcommand{\footrulewidth}{0pt}
\thispagestyle{fancy}

\section{Introduction}
\label{sec:intro}

Mixture-of-Experts (MoE) models increase capacity by replacing each dense feed-forward network 
with multiple experts while activating only a few per token~\cite{jacobs_adaptive_1991, mixtral, 
yang_qwen2_2024, dai_deepseekmoe_2024, deepseek-ai_deepseek-v2_2024}. This sparse activation limits per-token computation and parameter traffic, 
making MoE models attractive for local inference on PCs~\cite{mixtral_offloading, song2024promoe, ktransformers}.
Yet the complete set of expert weights often exceeds GPU memory.
Recent systems therefore keep all experts in host memory and execute activated experts 
across the CPU and GPU in parallel~\cite{llamacpp, ktransformers, hybrimoe, kamahori2024fiddler, LayerScope}. 
Such hybrid execution reduces expert-loading overhead by using CPU compute, but 
its performance hinges on balancing the GPU path, including GPU computation 
and expert-loading stalls, with the typically slower CPU path~\cite{hybrimoe}.

Unlike datacenter servers that dedicate resources to LLM inference~\cite{orca, vllm, 
zhong2024distserve}, PCs routinely run inference alongside other CPU applications. 
Such colocation is increasingly common in agentic workloads: 
coding agents interleave inference with CPU-intensive tools~\cite{codex, jimenez2024swebench, 
yang2024sweagentagentcomputerinterfacesenable, zhang2024autocoderoverautonomousprogramimprovement}, 
while multi-agent systems may run tools and inference concurrently~\cite{codex, 
zhang2025optimizing, fourney2024magentic, ccalikyilmaz2025optima, liu2026multi}. 
The inference engine generally has no prior knowledge of the colocated applications, 
their arrival times, execution durations, or scheduling priorities. 
Consequently, OS scheduling and contention for shared resources cause 
the effective CPU capacity available to inference to fluctuate rapidly and unpredictably, 
both across and within requests.
Therefore, the amount of expert work that the CPU can sustain cannot be determined reliably 
before execution.

\begin{figure}[t]
\begin{minipage}{1.\linewidth}
\centering
\includegraphics[width=\linewidth]{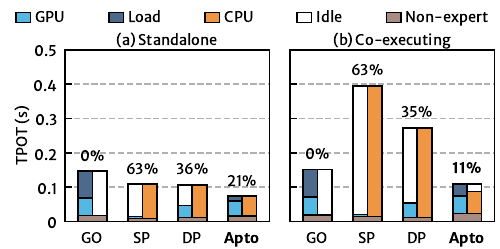}
\end{minipage} \begin{minipage}{1.\linewidth}
\caption{\small{\emph{
Decode latency (TPOT) breakdown for Qwen3-30B-A3B at batch size 4 when
running (a) standalone and (b) colocated with CPU-intensive workloads (e.g., parallel compilation).
GO, SP, and DP denote FineMoE~\cite{yu2026taming} (GPU-only), 
KTransformers~\cite{ktransformers} (static partitioning) and 
HybriMoE~\cite{hybrimoe} (dynamic partitioning), respectively.
Labels above the bars show the fraction of activated experts executed on the CPU.
}}}
\label{fig:intro:intro-breakdown}
\end{minipage}  \\[-5pt]
\end{figure}

Existing MoE inference systems are self-centered: they optimize expert execution 
assuming exclusive access to the PC's CPU, GPU, and memory resources. 
In GPU-only offloading, all activated experts execute on the GPU, making expert loading 
from host memory into GPU memory a major bottleneck~\cite{yu2026taming, song2024promoe, 
moe-infinity, mixtral_offloading}. 
Hybrid systems reduce this overhead by executing some experts on the CPU. 
Static partitioning (SP) avoids expert loading by letting the GPU execute only resident experts,
while the CPU executes the rest~\cite{kamahori2024fiddler, ktransformers}. 
Dynamic partitioning (DP) uses predicted execution times to determine
the CPU/GPU partition before expert execution begins
and overlaps expert loading with computation~\cite{hybrimoe,LayerScope,
yang2026freetoken,cao2025moe}.
With exclusive access to the CPU, they can eliminate or hide expert-loading stalls 
and improve inference performance, as shown in Figure~\ref{fig:intro:intro-breakdown}(a).

On shared PCs, however, this self-centered assumption does not hold. As shown in Figure~\ref{fig:intro:intro-breakdown}(b), when inference is colocated with
a CPU-intensive workload, static and dynamic partitioning still eliminate or
hide expert-loading time, yet their latencies increase to 3.64$\times$ and 2.58$\times$
their standalone values and reach 2.60$\times$ and 1.80$\times$
the GPU-only latency, respectively.
Because these methods continue to partition experts as if the full CPU were available, 
they assign nearly the same fraction of experts to the CPU despite contention. 
The CPU path consequently becomes the bottleneck, leaving the GPU underutilized or idle. 
A partition that better exploits the CPU under exclusive execution may therefore 
suffer greater degradation when resources are shared. 
The fundamental problem is not the quality of the partition under exclusive execution, 
but the assumption that this partition remains appropriate under colocation.

\stitle{Key insight.} 
Because CPU resources under colocation are dynamic and unpredictable, 
MoE inference should shift from \textbf{self-centered} upfront partitioning 
to \textbf{self-adaptive} runtime scheduling. 
Rather than predicting an optimal partition and committing all expert assignments 
before execution, the system should defer assignment decisions until execution and 
adapt the CPU/GPU work split as execution progresses. 
In this way, resource variation becomes feedback that guides scheduling rather than 
uncertainty that must be predicted.

\stitle{Our approach.}
To realize this principle, we design {\sys}, a self-adaptive CPU/GPU hybrid inference system 
that applies \textbf{dynamic scheduling}~\cite{polychronopoulos1987guided,chen2010tmr} 
to expert execution. 
{\sys} treats each activated expert as a task and lets the CPU and GPU executors claim new tasks 
only after completing their current ones. 
Task completion thereby feeds the realized capacity of each path directly into subsequent assignments, 
implicitly capturing interference from OS scheduling and shared resources.
When contention slows the CPU path, the CPU completes and claims fewer subsequent tasks, 
while the GPU naturally takes over more work. 
The CPU/GPU work split thus emerges from realized execution progress rather than being decided upfront. 
Although CPU contention is our primary focus, the mechanism also adapts to slowdowns 
on the GPU path.

Although dynamic scheduling provides flexibility,
general-purpose implementations introduce runtime overhead for task management
and coordination among multiple competing executors~\cite{vandierendonck2011unified,alvarez2021advanced,augonnet2011starpu,chase2005dynamic}.
However, several characteristics of MoE hybrid inference simplify these requirements
and enable efficient dynamic scheduling.
First, expert computations within an MoE block are independent until their outputs are merged,
avoiding the dependency-tracking overhead of general task runtimes~\cite{vandierendonck2011unified}.
Second, the set of expert tasks can be determined upfront,
without the need to handle dynamic task generation~\cite{chase2005dynamic}.
Third, CPU/GPU hybrid scheduling involves only two logical executors,
one for the CPU path and one for the GPU path,
avoiding complex coordination across large worker pools~\cite{prokopec2013near,alvarez2021advanced}.
A simple shared deque suffices to distribute expert tasks between the CPU and GPU paths.

\etitle{Affinity-aware scheduling.}
Dynamic scheduling adapts how much work each path receives,
but efficient execution also depends on which tasks it selects.
Expert affinities to the CPU and GPU are largely complementary:
GPU-resident experts, experts with more input tokens,
and frequently activated experts favor GPU execution,
whereas their counterparts favor CPU execution~\cite{kamahori2024fiddler,hybrimoe}.
{\sys} captures these preferences in a unified affinity ordering
that places the most GPU-preferred and CPU-preferred experts at opposite ends of the queue.
Each executor therefore claims its preferred remaining task from its own end,
preserving locality and processor efficiency without cross-executor coordination.

\etitle{Cancellable tail racing.}
Dynamic scheduling can rebalance only tasks that remain in the queue. 
Once the final expert has been claimed, a slowdown on its execution path can leave 
the other path idle and delay the entire MoE block. 
Since expert computation can be repeated,
{\sys} allows both executors to race the final expert 
when the two queue ends meet and accepts only the first completed result,
following the principle of backup execution~\cite{dean2004mapreduce}.
The two copies share the same PC resources,
so the slower copy can continue consuming resources or delay subsequent inference
even after its result is no longer needed.
{\sys} therefore cancels the slower copy's remaining execution once the faster copy completes,
limiting both tail latency and unnecessary resource consumption.

We implemented {\sys} atop llama.cpp~\cite{llamacpp}
with ProMoE~\cite{song2024promoe} for expert caching and prefetching,
and evaluate it against state-of-the-art GPU-only and CPU/GPU hybrid baselines.
When the inference process exclusively occupied the PC resources, 
{\sys} achieved average speedups of 2.96$\times$ and 1.52$\times$ in the prefill and decode stages, respectively. 
When co-executing with a CPU-intensive workload under different configurations,
{\sys} achieved average speedups of 5.97$\times$ and 3.45$\times$ in the prefill and decode stages, respectively,
by adapting to dynamic CPU availability.
Meanwhile, the colocated workload achieved 1.08$\times$ the throughput obtained alongside the baselines on average.
For real-world coding agent applications, {\sys} achieved an average speedup of 2.02$\times$ over the baselines.
A controlled experiment further confirms that {\sys} also adapts to GPU contention.

\section{Background and Motivation}
\label{sec:bg}

\subsection{CPU/GPU hybrid inference for MoE-based LLMs}

Compared with cloud deployment, 
deploying large language models (LLMs) on personal computers (PCs) offers improved privacy, 
offline availability, and cost efficiency~\cite{xu2024device, qu2025mobile, edgemoe, song_powerinfer_2023}.
Efficient inference on these devices requires careful use of limited compute and memory bandwidth.
The Mixture-of-Experts (MoE) architecture addresses these 
constraints by dividing the FFN into multiple experts and activating only a subset for each token.
In each MoE block, a gating function selects the experts to process each token, 
and their outputs are combined using a weighted sum.
This sparse activation reduces computation and memory access, 
making MoE-based LLMs well-suited for local inference on PCs with consumer-grade GPUs.

However, the large number of experts increases the memory footprint, 
often exceeding the limited memory capacity of consumer-grade GPUs. 
CPU/GPU hybrid execution~\cite{llamacpp, ktransformers, hybrimoe, kamahori2024fiddler, LayerScope} 
offloads expert parameters to host memory, keeps only a subset of experts in GPU memory, 
and computes activated experts using both CPU and GPU resources. 
This design reduces GPU memory requirements and mitigates performance degradation 
by reducing CPU-GPU data transfer and exploiting the CPU for parallel computation. 
The performance of CPU/GPU hybrid execution is largely determined by load balancing between the GPU path, 
which includes GPU computation and waiting for parameter transfer, and the CPU path~\cite{hybrimoe}.

\subsection{LLM Inference on Shared PCs}

\nostitle{Colocation of LLM inference and other applications.}
Unlike cloud servers, where resources are typically dedicated to LLM inference,
PCs commonly run inference alongside other applications.
These applications can use shared resources, especially CPU resources,
creating contention that affects CPU/GPU hybrid inference.

Moreover, emerging LLM applications increasingly require coordinated execution
between inference and external tools, making such colocation more common.
For example, coding agents~\cite{codex, yang2024sweagentagentcomputerinterfacesenable,
zhang2024autocoderoverautonomousprogramimprovement}
complete tasks by repeatedly requesting LLM responses and executing CPU-based tools.
Multi-agent systems~\cite{codex, zhang2025optimizing, fourney2024magentic,
ccalikyilmaz2025optima, liu2026multi}
run multiple agents in parallel to complete complex tasks or improve efficiency.
CPU-based tools and inference requests from different agents can overlap,
resulting in concurrent tool execution and LLM inference on the same PC.

\stitle{Dynamic resource contention.}
Contention changes as colocated applications start, finish, or vary their resource demands.
System scheduling decisions further affect how resources
are shared between inference and other applications.
The inference engine generally has no advance knowledge of these changes.
As contention varies, so does the effective CPU capacity available to inference,
causing CPU-side expert execution times to fluctuate.

To characterize the impact of dynamic CPU contention in real-world agent workloads,
we collect execution traces of GPT-5.5-powered SWE-agent~\cite{yang2024sweagentagentcomputerinterfacesenable} on seven SWE-bench tasks~\cite{jimenez2024swebench}, 
which resolve real-world GitHub issues across three Java or Rust repositories.
When replaying these traces, we pause each agent worker according to the recorded LLM response waiting time, 
preserving the temporal structure of the original agent execution.

\begin{figure}[t]
\centering
\includegraphics[width=\linewidth]{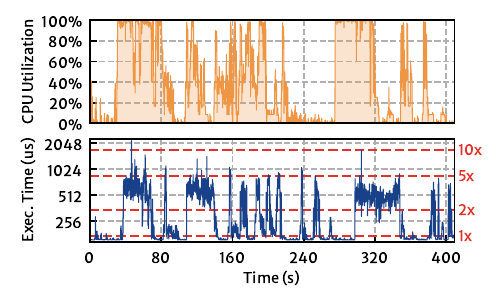}
\caption{\small{\emph{
Timeline of CPU utilization when replaying coding-agent traces with four concurrent agent workers~(a),
and the corresponding CPU execution time of one expert in Qwen3-30B-A3B~\cite{qwen3technicalreport}~(b).
Dashed lines in~(b) indicate slowdown levels relative to the expert's uncontended CPU execution time, as labeled.
}}}
\label{fig:bg:agent-trace-replay-motiv}
\end{figure}

Figure~\ref{fig:bg:agent-trace-replay-motiv}(a) shows CPU utilization
during trace replay with four concurrent agent workers.
Bursty agent tool calls cause utilization to alternate repeatedly
between low-utilization periods and brief high-utilization bursts.
In particular, the CPU utilization of agent tools increases
from 2.2\% to 96.1\% within 0.9 seconds,
illustrating how rapidly their CPU demand can change.
Figure~\ref{fig:bg:agent-trace-replay-motiv}(b) shows the corresponding variation
in the CPU execution time of one expert in Qwen3-30B-A3B~\cite{qwen3technicalreport}.
Concurrent tool execution causes substantial fluctuations in expert execution time,
with P75 and P99 slowdowns of 3.0$\times$ and 5.7$\times$,
respectively, compared with the uncontended case.

\stitle{Need for adaptive scheduling.}
Changes in CPU performance can invalidate a previously balanced CPU/GPU expert allocation.
To maintain load balance, the inference engine must adapt expert assignments
to runtime execution performance.
When CPU contention slows expert execution, more work should shift to the GPU.
The allocation should continue to adapt as contention changes.
The same principle also applies when GPU contention changes
the relative performance of the two paths.

\begin{figure*}[t!]
\centering
\begin{minipage}{\linewidth}
\centering
\includegraphics[width=\linewidth]{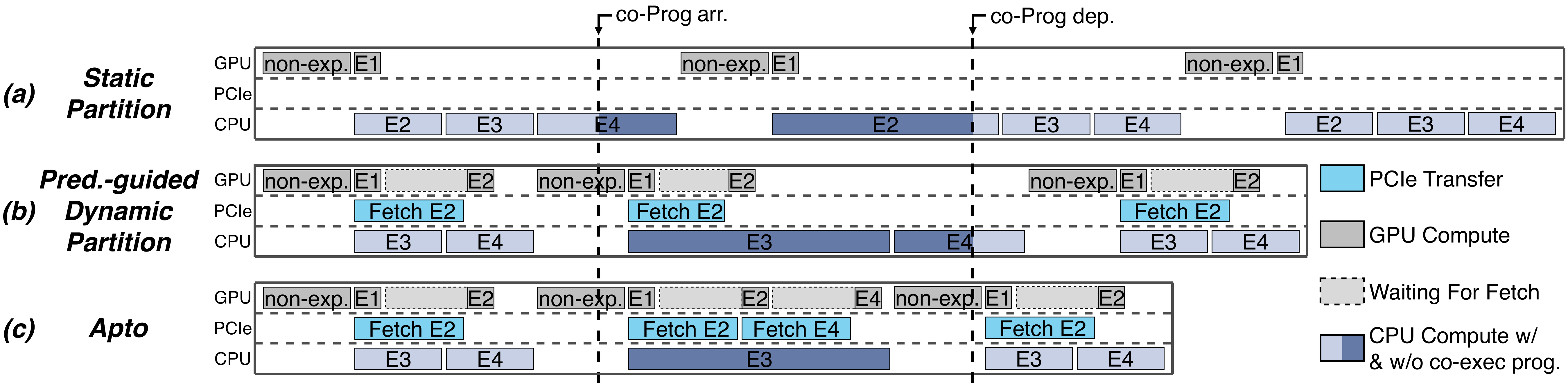}
\end{minipage} \\[2pt]
\begin{minipage}{1\linewidth}
\caption{\small{\emph{
An example of expert execution patterns under different CPU/GPU hybrid methods across three transformer layers, 
each containing four activated experts: one GPU-resident expert (E1) and three offloaded experts (E2-E4).
The illustrated time slice contains one co-execution CPU program (co-Prog) arriving (arr.) and departing (dep.).
}}}
\label{fig:exist-works}
\end{minipage}
\end{figure*}

\subsection{Existing CPU/GPU Hybrid Methods}

Existing CPU/GPU hybrid methods partition expert workloads before expert execution begins, 
using either static or dynamic partitioning, as illustrated in Figure~\ref{fig:exist-works}.
These methods are self-centered: they optimize the partition assuming exclusive access to the PC's CPU, GPU, 
and memory resources, without adequately accounting for runtime contention from colocated applications.

\stitle{Static partitioning.}
Recent works~\cite{kamahori2024fiddler, ktransformers} propose static partitioning,
which fixes the CPU/GPU partition during initialization by placing frequently used experts in GPU memory
and offloading the remaining experts to CPU memory.
The experts are executed on their assigned processors,
without any CPU--GPU expert parameter transfer during inference,
as illustrated in Figure~\ref{fig:exist-works}(a).
Since the GPU provides higher computational capacity than the CPU,
this strategy aims to assign more expert executions to the GPU and fewer to the CPU,
thereby improving load balance.

However, modern MoE models often exhibit nearly uniform expert access patterns~\cite{song2024promoe},
so the fraction of activated experts served by the GPU is largely determined by GPU memory capacity.
Limited GPU memory can therefore leave the CPU path overloaded,
while the fixed partition prevents the GPU from taking over its work.
Even during standalone inference, the GPU remains idle for 93.7\% of the expert execution time 
while waiting for CPU-assigned experts to complete (Figure~\ref{fig:intro:intro-breakdown}(a)).
CPU contention further prolongs this wait, increasing the idle fraction
to 98.4\% (Figure~\ref{fig:intro:intro-breakdown}(b)).

\stitle{Dynamic partitioning.}
To further improve load balance, recent works~\cite{hybrimoe, LayerScope, yang2026freetoken, cao2025moe}
adopt dynamic partitioning. 
It assigns activated experts to the CPU or GPU at runtime before execution,
allowing experts stored in host memory to be transferred to the GPU for execution.
This flexibility allows the inference engine to shift part of 
the CPU workload to the GPU to improve load balance, as shown in Figure~\ref{fig:exist-works}(b).
Existing methods use performance models built through offline profiling
to predict expert execution times on the CPU and GPU paths
and determine the CPU/GPU partition based on the predicted execution times.

However, these prediction-guided methods assume that runtime CPU performance matches offline profiling,
which does not always hold during co-execution.
When colocated CPU programs slow down expert execution,
the scheduler still relies on estimates of uncontended CPU performance
and consequently underestimates CPU execution time.
For example, Figure~\ref{fig:intro:intro-breakdown} shows that
HybriMoE assigns nearly the same fraction of activated experts
to the CPU during standalone inference and co-execution
(35.5\% and 35.4\%, respectively).
This allocation fails to reflect the CPU performance degradation,
increasing GPU idle time to 3.73$\times$ its standalone value.

Moreover, performance models can be inaccurate even during standalone inference.
For example, HybriMoE's performance model~\cite{hybrimoe}
exhibits a mean absolute relative error of 71\% on Qwen3-30B-A3B with batch size 4
due to the complex behavior of highly optimized CPU and GPU operators.
These prediction errors lead to suboptimal CPU/GPU partitions,
leaving the actual execution times of the two paths imbalanced.
As shown in Figure~\ref{fig:intro:intro-breakdown}(a),
the GPU still spends 62.4\% of expert execution time idle waiting for the CPU, even without contention.

Finally, each performance model is specific to a hardware platform,
MoE model, and system configuration (e.g., the number of CPU executor threads).
Changes to the deployment environment require rebuilding or updating the performance model,
adding profiling and maintenance overhead that limits the practicality
of prediction-guided dynamic partitioning in real-world PC environments.

\section{Approach and Overview}

This paper presents {\sys}, a self-adaptive CPU/GPU hybrid inference system for MoE-based LLMs on shared PCs.
As shown in Figure~\ref{fig:exist-works}(c),
when CPU contention slows down CPU-side execution,
{\sys} reduces the fraction of experts dispatched to the CPU and shifts more work to the GPU.
As CPU performance recovers, it increases the fraction of experts
executed on the CPU to utilize the available resources.
This allows {\sys} to maintain load balance and adapt to the 
resource contention and OS scheduling decisions on shared PCs.

\stitle{Our approach: dynamic scheduling.}
Rather than predicting execution times and partitioning experts
between the CPU and GPU before execution begins,
{\sys} assigns experts dynamically as execution progresses.
Each activated expert is treated as a task, and each executor claims its next task after completing the current one.
Task completion thus provides direct feedback for subsequent assignments,
implicitly reflecting the capacity available to each execution path.
When contention slows one path, it completes and claims fewer tasks,
allowing the other path to handle more of the remaining work.
The CPU/GPU work split therefore adapts to actual execution progress
without requiring an upfront partition.

MoE hybrid inference simplifies the task management and coordination
requirements of general-purpose dynamic scheduling.
First, experts within an MoE block execute independently before their outputs are merged,
so scheduling them does not require tracking inter-task dependencies~\cite{vandierendonck2011unified}.
Second, the gate determines the task set before expert execution begins.
The queue is initialized once per block and receives no new expert tasks during execution,
avoiding dynamic task insertion and queue growth~\cite{chase2005dynamic}.
Third, task assignment involves only two logical executors,
one for the CPU path and one for the GPU path.
A shared double-ended queue (deque) therefore suffices without the coordination mechanisms
needed to scale scheduling across large worker pools~\cite{prokopec2013near,alvarez2021advanced}.
These characteristics make dynamic scheduling a natural fit
for CPU/GPU hybrid MoE inference.

\begin{figure}[t]
\centering
\begin{minipage}{\linewidth}
\centering
\includegraphics[width=\linewidth]{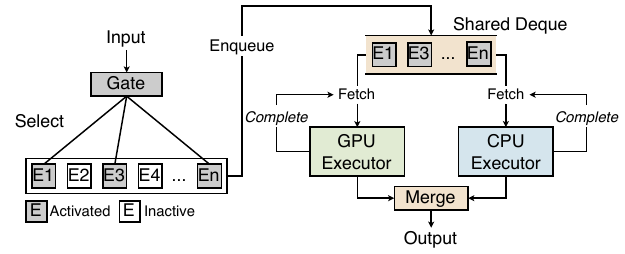}
\end{minipage} \\[2pt]
\begin{minipage}{1\linewidth}
\caption{\small{\emph{
The workflow of {\sys} for scheduling expert tasks between CPU and GPU executors 
to compute the activated experts in an MoE block.
}}}
\label{fig:overview:sys-arch}
\end{minipage}
\end{figure}

\stitle{Overview of {\sys}.}
During initialization, {\sys} places the non-expert components of the LLM,
including self-attention blocks and MoE gate functions, in GPU memory.
All experts are stored in host memory, with a subset cached in GPU memory.
During inference, the non-expert components execute on the GPU,
while activated experts are scheduled to run on either the CPU or the GPU.

Figure~\ref{fig:overview:sys-arch} illustrates the execution workflow of an MoE block.
The gate function selects the activated experts, which are enqueued as tasks in a shared deque.
The GPU and CPU executors draw tasks from opposite ends and execute them in parallel.
Each executor follows a fetch--execute loop,
fetching its next task after completing the current one.
The GPU executor uses cached expert parameters when available
and loads missing parameters from host memory,
while the CPU executor accesses parameters directly in host memory.
Once all activated experts have completed execution,
their outputs are merged on the GPU to produce the block's output.

\vspace{-4pt}
\section{Dynamic Scheduling}
\label{sec:hybrid}

\subsection{Affinity-aware Scheduling}

Dynamic scheduling should consider task affinity to execution paths,
as task assignments affect both data movement and computational efficiency.
Existing dynamic scheduling techniques consider affinity using factors
such as data locality and cache reuse~\cite{markatos1994using}, data transfer costs~\cite{augonnet2010data},
and relative execution speeds across heterogeneous processors~\cite{augonnet2011starpu}.
This motivates the need for {\sys} to consider
expert tasks' affinity to the CPU and GPU execution paths in hybrid MoE inference
to support \emph{affinity-aware scheduling}.

\begin{figure}[t]
\centering
\begin{minipage}{\linewidth}
\centering
\includegraphics[width=1\linewidth]{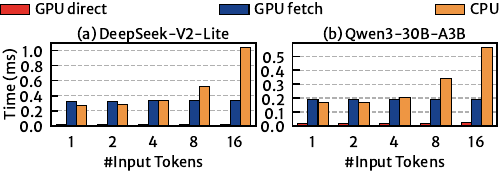}
\end{minipage} \\[2pt]
\begin{minipage}{1\linewidth}
\caption{\small{\emph{
Expert execution time of DeepSeek-V2-Lite (a) and Qwen3-30B-A3B (b)
on CPU and GPU across different input token counts.
GPU direct and GPU fetch indicate GPU execution with cached experts
and GPU execution after fetching experts from CPU memory, respectively.
}}}
\label{fig:hybrid:expert-time}
\end{minipage}
\end{figure}

In hybrid MoE inference, {\sys} considers three factors that shape
expert tasks' execution-path affinity: cache status, input token count, and activation frequency.
First, cache status determines whether GPU execution requires loading expert weights from host memory.
Cache hits avoid this transfer, while cache misses increase average 
GPU-path execution time to 13.6$\times$ that of cache hits, as shown in Figure~\ref{fig:hybrid:expert-time}.
This favors prioritizing cache-hit experts on the GPU path.

Second, input token count affects the relative computational efficiency of CPU and GPU execution.
In each forward pass, multiple input tokens can be processed, and each token selects experts independently,
leading to varying input token counts across experts.
Figure~\ref{fig:hybrid:expert-time} shows that as the input token count increases, CPU execution time grows significantly, 
while the execution time of both GPU execution modes remains relatively stable. 
This is because GPUs can exploit parallelism across tokens to handle the additional computation efficiently.
This favors prioritizing experts with more input tokens on the GPU path
and those with fewer input tokens on the CPU path.

Third, activation frequency affects the potential benefit of
loading an expert into the GPU cache for reuse in future iterations~\cite{ktransformers}.
If a frequently activated expert is not cached on the GPU, loading it into
the GPU cache can improve future cache hit rates and reduce execution time.
This favors prioritizing experts with higher activation frequencies on the GPU path.

Prior affinity-aware dynamic scheduling techniques encode task preferences
through queue organization~\cite{markatos1994using,vet2012multigrain}.
MoE hybrid inference involves only two logical executors, whose preferences
under the three factors above are largely complementary.
This allows {\sys} to represent both executors' preferences in a unified affinity ordering.
Since the gate identifies all activated experts before execution,
{\sys} constructs this ordering once per MoE block before enqueuing tasks,
and places the most GPU-preferred and CPU-preferred experts at opposite ends of the shared deque.
Each executor then claims its preferred remaining task from its own end.
The ordering guides task choice, while the CPU/GPU work split emerges from actual execution progress.

To construct this ordering, {\sys} assigns different priorities to the three factors.
Cache status and input token count directly affect execution time in the current MoE block, 
while activation frequency only offers potential benefits in future iterations.
{\sys} therefore gives the first two factors precedence over activation frequency.
Among them, {\sys} prioritizes cache status, guided by the observation
that cached experts execute substantially faster on the GPU than on the CPU,
even at small token counts (Figure~\ref{fig:hybrid:expert-time}).
The resulting priority hierarchy uses cache status as the primary factor,
input token count as the secondary factor, and activation frequency
as the tertiary factor.

Before enqueueing, {\sys} records each expert task's cache status
and input token count, and obtains its activation frequency estimate
from offline profiling and online incremental updates.
Each task $e$ is assigned a priority key
\[
K(e) = \big(c_e, n_e, f_e\big),
\]
where $c_e \in \{0,1\}$ indicates GPU cache residency when the ordering is constructed,
$n_e$ is the input token count, and $f_e$ is the estimated activation frequency.
Tasks are enqueued in descending lexicographic order of $K(e)$.
The GPU executor claims tasks from the front, while the CPU executor
claims tasks from the back.

\subsection{Cancellable Tail Racing}

\begin{figure}[t]
\centering
\begin{minipage}{\linewidth}
\centering
\includegraphics[width=1\linewidth]{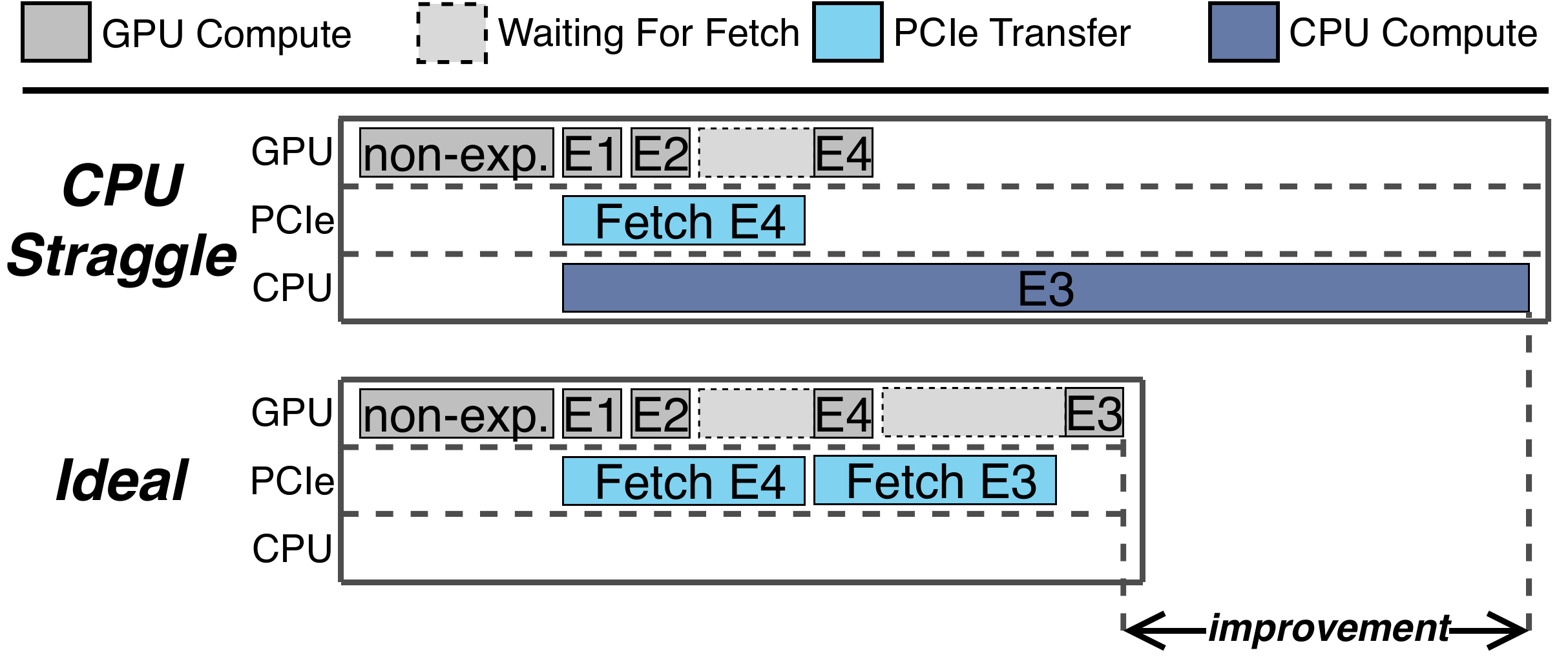}
\end{minipage} \\[2pt]
\begin{minipage}{1\linewidth}
\caption{\small{\emph{
An example of the straggler problem in dynamic scheduling, where the CPU executor becomes the straggler. 
Experts 1 and 2 are cache-hit experts, while experts 3 and 4 are cache-miss experts.
}}}
\label{fig:straggler}
\end{minipage}
\end{figure}

Dynamic scheduling adapts task assignments as execution progresses, 
but it can only rebalance tasks that remain in the queue.
In common queue-based scheduling schemes, tasks are dequeued before execution~\cite{markatos1994using,vet2012multigrain, blumofe1999scheduling}.
Once the queue is empty, the executor processing the final unfinished expert can become a straggler 
while the other executor, which can complete the same task faster, is already idle.
The MoE block must then wait for the straggler to finish,
despite the availability of a faster execution path.
Figure~\ref{fig:straggler} illustrates this problem:
the CPU executor is still executing expert 3 after the GPU executor has completed experts 1, 2, and 4.
In this example, executing all four experts on the GPU would achieve a shorter overall completion time.

The straggler problem can arise when the execution time of an expert task differs
significantly between the two executors, which can occur in several MoE hybrid inference scenarios.
Contention from co-executing processes can slow CPU execution, making the CPU executor the straggler.
When all activated experts are GPU cache hits, the GPU can execute them much faster than the CPU, 
also making the CPU executor the straggler.
Limited CPU--GPU bandwidth or PCIe contention can instead delay
the transfer of a cache-miss expert, making the GPU executor the straggler.

Backup execution is a common approach to mitigating stragglers in
parallel systems~\cite{dean2004mapreduce,zaharia2008improving,ananthanarayanan2013effective}.
It launches an additional copy of an unfinished task on an idle executor and accepts the first completed result,
allowing the faster executor to finish the task originally assigned to the straggler.
Inspired by backup execution, {\sys} introduces \emph{tail racing}:
the CPU and GPU executors race to complete the final unfinished expert,
and only the first completed result is used.
Making tail racing efficient requires a low-overhead trigger
and a mechanism to limit redundant work.

Expert computation in MoE inference is repeatable, enabling backup execution
provided that only one copy's result is used in the output merge.
Moreover, the two logical executors consume tasks from opposite ends
of the deque, so their convergence on the same expert provides
a natural trigger for backup execution.
By keeping an expert task in the queue until it completes, 
{\sys} allows the idle executor to reach the same task and race the other executor to complete it.
This enables an efficient implementation of tail racing.

In practice, {\sys} implements tail racing by deferring task removal until execution completes.
Each executor first peeks at the task at its preferred end of the deque and executes it without removing it.
After execution, the executor atomically attempts to dequeue the task,
succeeding only if it has not already been removed.
When both executors execute the same expert, only one can successfully dequeue the task and mark it as completed.
Only the result from that executor is used in the output merge,
ensuring that the expert contributes exactly once.

The two copies of the final expert task can contend for host-memory bandwidth
when CPU computation overlaps with GPU parameter loading over PCIe.
On our platform, these operations individually achieve 51.7 and 51.4 GiB/s,
respectively, with their sum only slightly exceeding the measured
host-memory bandwidth of 98 GiB/s.
This suggests that tail racing introduces limited bandwidth contention during standalone inference.
However, co-execution with bandwidth-intensive programs reduces
the bandwidth available to inference, potentially increasing contention and the overhead of tail racing.

The two copies share the same PC resources, and the losing copy can
continue consuming resources or delay subsequent inference even after its result is no longer needed.
{\sys} therefore cancels its remaining execution to limit unnecessary resource consumption.
To enable fast cancellation, {\sys} splits each expert task into chunks, with a completion check at each chunk boundary.
Once all expert tasks in the MoE block have completed, the executor skips the remaining chunks.
Experts often consist of several linear layers~\cite{qw1, deepseek-ai_deepseek-v2_2024, qwen3technicalreport},
which provide natural boundaries for chunking computation and weight transfers.
The implementation details of CPU and GPU cancellation are described in \S\ref{sec:impl}.

\section{Detailed Design and Implementation}
\label{sec:impl}

\SetKwFunction{Peek}{peek}
\SetKwFunction{Dequeue}{dequeue}

\subsection{GPU Executor}
\label{sec:gpu-executor}

\begin{algorithm}[t]
\small
\newcommand{\GreenCommentSty}[1]{\textcolor{green!50!black}{\texttt{#1}}}
\SetCommentSty{GreenCommentSty}
\caption{GPU execution worker}
\label{alg:gpu-worker}
\KwData{$Q$: Shared Deque; $N$: Unfinished Task Count}
\KwData{$M$: Maximum In-flight GPU Tasks}
\SetKw{Break}{break}
$P \gets [\,]$ \tcp*{In-flight FIFO}
\While{$N > 0 \lor P \neq [\,]$}{
    \While{$|P| < M \land N > 0$}{
        \tcp{Ready: cached and not in flight}
        $task \gets Q.\texttt{peek\_front}(\mathrm{ready})$\;
        \lIf{$task = \varnothing$}{\Break}
        $\texttt{launch\_gpu\_kernel}(task)$\;
        $P.\texttt{push}(task)$\;
    }
    \If{$P \neq [\,]$}{
        $\texttt{wait\_for\_completion}(P.\texttt{front}())$\;
        \If{$Q.\texttt{dequeue}(P.\texttt{pop}())$}{
            $N \gets N - 1$\;
        }
    }
}
\end{algorithm}

The GPU executor executes experts whose parameters can be either cached or uncached in GPU memory.
Overlapping the transfer of cache-miss expert parameters with the execution of cache-hit experts 
is crucial for improving GPU execution efficiency~\cite{yu2026taming, song2024promoe, moe-infinity, mixtral_offloading}.
To enable this overlap, the GPU executor uses two worker threads:
a GPU execution thread that launches GPU kernels for cache-hit experts, 
and a fetcher thread that transfers expert parameters from host memory to GPU memory.

\stitle{Peek.}
The \Peek{} operation is used to get the next task in the shared deque without removing it,
which is essential to enable tail racing.
Both threads perform \Peek{} on the shared deque,
searching from the front according to their respective roles.
The GPU execution thread peeks cache-hit expert tasks,
while the fetcher thread peeks cache-miss expert tasks.
Once an expert's parameters have been fully transferred to GPU memory,
the fetcher thread marks the task as a cache hit,
making it available to the GPU execution thread through \Peek{}.
This allows GPU computation to overlap with data transfer.

\stitle{Dequeue.}
The GPU execution thread performs \Dequeue{} after an expert's GPU kernels complete.
Since GPU kernels run asynchronously, waiting immediately after each launch would leave the GPU idle 
during \Dequeue{} and the next kernel launch.
To overlap this control flow with GPU computation,
{\sys} allows a limited number of in-flight expert tasks,
i.e., tasks submitted to the GPU whose completion has not yet been processed,
and tracks their completion with events.

As shown in Algorithm~\ref{alg:gpu-worker},
the GPU execution thread continues to peek and launch cache-hit expert tasks,
skipping those already in flight, until the limit is reached or no eligible task remains.
It then waits for the oldest in-flight task to complete and performs \Dequeue{}.
A successful \Dequeue{} decrements a shared counter that tracks unfinished expert tasks
across both the CPU and GPU executors.
The thread repeats this process, allowing completion handling to overlap
with the GPU execution of subsequent experts.
Our implementation sets the in-flight limit to two.
When the counter reaches zero, all required expert tasks have completed,
and the GPU execution thread exits its execution loop after processing any remaining in-flight tasks.

\stitle{Cancellation.}
{\sys} uses the unfinished task counter to detect when ongoing redundant execution is no longer needed.
Since GPU kernels run asynchronously and their execution time is negligible
compared with parameter transfer time,
{\sys} only cancels redundant transfers for cache-miss experts
while allowing already-launched kernels to finish.
For on-demand fetches, the fetcher thread transfers parameters in chunks
and synchronizes after each chunk to check whether the unfinished task counter has reached zero.
If so, it stops transferring the remaining chunks.

\stitle{Cache management.}
{\sys} integrates ProMoE~\cite{song2024promoe} for expert cache management, enabling expert prefetching
that overlaps data transfer with non-expert computation to improve inference performance.
To keep prediction efficient, {\sys} runs a lightweight structure predictor on the GPU to avoid CPU contention
and hide its computation behind CPU execution and cache-miss expert transfers.
The predicted expert tasks are placed in a prefetch queue,
which the fetcher thread processes at lower priority than on-demand transfers.

\subsection{CPU Executor}

\begin{algorithm}[t]
\small
\newcommand{\GreenCommentSty}[1]{\textcolor{green!50!black}{\texttt{#1}}}
\SetCommentSty{GreenCommentSty}
\caption{CPU execution worker}
\label{alg:cpu-worker}
\KwData{$Q$: Shared Deque; $N$: Unfinished Task Count}
\KwData{Shared: $task$, $cancel$, $\texttt{barrier}$}
\KwData{Thread-local: $i$ (thread ID)}
\SetKw{Break}{break}
\SetKw{False}{false}
\While{true}{
    \If{$i = 0$}{
        $task \gets Q.\texttt{peek\_back}()$\;
        $cancel \gets \False$\;
    }
    $\texttt{barrier.wait}()$\;
    \lIf{$task = \varnothing$}{\Break}

    \tcp{Cooperative execution in chunks}
    \ForEach{$chunk \in task.\texttt{chunks}$}{
        $\texttt{execute\_chunk}(i, chunk)$\;
        \lIf{$i = 0$}{$cancel \gets (N = 0)$}
        $\texttt{barrier.wait}()$\;
        \lIf{$cancel$}{\Break}
    }

    \If{$i = 0 \land \lnot cancel$}{
        \If{$Q.\texttt{dequeue}(task)$}{
            $N \gets N - 1$\;
        }
    }
}
\end{algorithm}

The CPU executor uses multiple worker threads to execute expert tasks cooperatively.
When entering an MoE block and starting hybrid execution, the inference process wakes all CPU workers,
which then follow Algorithm~\ref{alg:cpu-worker}.
The first worker acts as the master: it performs \Peek{} from the back of the shared deque to fetch the next expert task,
then synchronizes with the others through a barrier so that all workers share the same task metadata.
The workers execute the task in parallel, chunk by chunk.
After each chunk, the master checks whether the unfinished task counter has reached zero.
If so, it sets a cancellation flag to stop the other workers from executing remaining chunks.
Otherwise, after the task completes, the master performs \Dequeue{} to remove it from the queue
and decrements the counter if successful.

\subsection{CUDA Graph Support}

CUDA graph~\cite{cudaProgrammingGuide} is commonly used to reduce kernel launch overhead and improve GPU efficiency.
Although {\sys}'s hybrid execution of MoE experts is dynamic and difficult to capture with a static CUDA graph,
non-expert components are executed on the GPU and follow static execution patterns, allowing them to benefit from CUDA graphs.
Specifically, {\sys} captures a subgraph for each non-expert component
and launches the corresponding graph at runtime.
For expert computation, the bottleneck lies in data transfer and CPU execution,
making kernel launch overhead negligible.

\subsection{Support for Unified Memory Architecture}

Recent AI PCs~\cite{intelcorporationIntelCoreUltra2025,appleincMacBookPro2025,
qualcommLaptopsPoweredSnapdragon} and mini AI supercomputers~\cite{beelinkBeelinkGTR9Pro2025,
appleincMacStudio2025} increasingly adopt 
Unified Memory Architecture (UMA)~\cite{power2013heterogeneous,yang2027libumsh},
in which the CPU and GPU share physical memory.
This allows all model parameters, including expert parameters,
to reside in the typically larger host memory and to be accessed directly by both processors.
LLM inference on these systems is typically bounded by memory bandwidth~\cite{chen2025characterizing, wei2025ghidorah},
which GPU-only execution can often saturate.
However, GPU contention from other applications can degrade inference performance.
Adaptive hybrid execution remains useful in this setting,
as it can dynamically shift expert computation to the CPU when the GPU is busy.
We therefore extend {\sys} to UMA-based PCs to retain this adaptability.

On UMA-based PCs, {\sys} reuses the shared deque and the CPU and GPU executors.
Since all expert parameters are directly accessible in shared memory,
{\sys} treats all experts as cache-hit experts and does not start the fetcher thread.
Accordingly, affinity-aware scheduling orders expert tasks solely by their input token counts,
and ignores cache status and activation frequency.

When the GPU is not contended by other applications,
concurrent CPU execution introduces competition for shared memory bandwidth,
degrading inference performance compared with GPU-only execution.
To avoid this, {\sys} modifies the CPU executor to enable hybrid execution
only when GPU progress stalls.
Before fetching a task, the CPU executor checks GPU task completion.
If no GPU expert task completes within a time window,
{\sys} treats this as an indication of GPU contention from other applications
and allows the CPU executor to fetch and execute an expert task.
This reduces unnecessary CPU--GPU memory-bandwidth contention
while preserving adaptability to GPU contention.

\section{Evaluation}
\label{sec:eval}

We implemented {\sys} as a MoE LLM inference engine on top of llama.cpp~\cite{llamacpp}.
At inference time, a custom operator received the gate logits and expert selections of each MoE block, 
triggering {\sys}'s core module to schedule hybrid execution.
Both CPU and GPU executors invoked llama.cpp operators to compute expert outputs.

\stitle{Hardware.}
We conducted our evaluation on a PC equipped with an NVIDIA RTX 5080 GPU
(16 GB GDDR7), an Intel Core Ultra 9 285K CPU with 8 performance cores
(8 threads) and 16 efficient cores (16 threads), and 96 GB of host DRAM.
The GPU was connected to the CPU via PCIe 5.0 $\times$ 16, providing a unidirectional bandwidth of 64 GB/s.
We restricted the evaluated workloads to 12 CPU cores (8 performance cores and 4 efficient cores).
We also evaluated {\sys} on NVIDIA DGX Spark~\cite{skende171NVIDIAGB102026}
to assess its performance on the UMA platform, as detailed in \S\ref{sec:eval:uma}.

\stitle{Workloads.}
For LLM inference, we used three MoE-based LLMs: Qwen1.5-MoE~\cite{qw1} (QW-1),
DeepSeek-V2-Lite~\cite{deepseek-ai_deepseek-v2_2024} (DS-2), and Qwen3-30B-A3B~\cite{qwen3technicalreport} (QW-3).
These models differed in both the number of experts activated per token
(4, 6, and 8, respectively) and their activation patterns.
We used batch sizes of 1 and 4 to further vary the patterns of expert tasks.
Prompts were drawn from ShareGPT~\cite{sharegpt} and each request generated 48 tokens.

For experiments involving co-execution of LLM inference and CPU programs,
we used compilation, which was a representative CPU workload commonly encountered in coding-agent scenarios.
Specifically, we used Ninja to compile zstd~\cite{zstd}.
For experiments involving co-execution of LLM inference and GPU programs,
we used ResNet-50~\cite{he2016deep} inference implemented in
PyTorch as the GPU workload.
Table~\ref{tab:workload-config} summarized the evaluation scenarios,
including standalone inference and co-execution with CPU or GPU workloads.

\begin{table}[t]
\centering
\caption{
\textit{Co-executing workload configurations across evaluation scenarios.
Compile/$n$ denoted $n$ parallel compilation jobs.
Priority was relative to LLM inference;
Pattern indicated continuous or intermittent execution.
Solo denoted standalone inference.}
}
\label{tab:workload-config}
\small
\begin{minipage}{\linewidth}
\begin{tabular*}{\linewidth}{@{\extracolsep{\fill}}lllr@{}}
\toprule
\textbf{Scenario} & \textbf{Workload} &
\textbf{Priority} & \textbf{Pattern} \\
\midrule
Solo    & --         & --    & --                 \\
Low     & Compile/2  & Equal & Continuous         \\
Medium  & Compile/6  & Equal & Continuous         \\
High    & Compile/10 & Equal & Continuous         \\
Dynamic & Compile/10 & Equal & 5\,s on / 5\,s off \\
HP      & Compile/6  & High    & Continuous         \\
LP      & Compile/6  & Low     & Continuous         \\
GPU     & ResNet-50  & Equal & Continuous         \\
\bottomrule
\end{tabular*}
\end{minipage}
\end{table}

\begin{figure*}[t]
\centering
\begin{minipage}{\linewidth}
\centering
\includegraphics[width=1\linewidth]{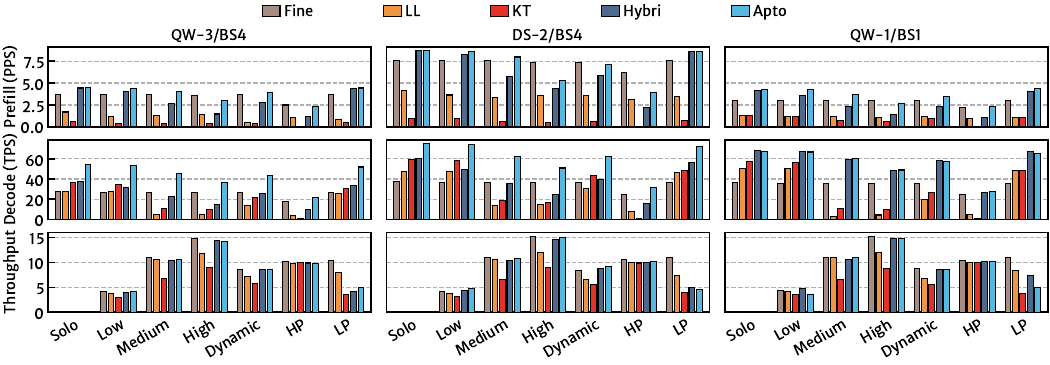}
\end{minipage} \\[2pt]
\begin{minipage}{1\linewidth}
\caption{\small{\emph{
Comparison of the inference performance and co-executing workload throughput
for {\sys} and its baselines when serving QW-3 and DS-2 with batch size 4
and QW-1 with batch size 1 across the Solo and CPU co-execution scenarios.
}}}
\label{fig:eval:overall}
\end{minipage}
\end{figure*}

\stitle{Baselines.}
To demonstrate the effectiveness of our design, we compared {\sys} against 
state-of-the-art MoE inference baselines designed for limited GPU memory.
(1) FineMoE~\cite{yu2026taming} (\textbf{Fine})
was a state-of-the-art GPU-only method that improved inference performance through expert prefetching.
It used fine-grained expert tracing to accurately predict future activated experts 
and overlap their data transfer with GPU computation.
(2) Llama.cpp~\cite{llamacpp} (\textbf{LL}) 
partitioned experts at the layer level, placing some layers' experts on the GPU
and offloading others to the CPU.
In the prefill stage, it transferred all offloaded experts to the GPU before execution.
In the decode stage, it executed each expert on its assigned processor.
It did not overlap GPU execution with CPU execution or data transfer.
(3) KTransformers~\cite{ktransformers} (\textbf{KT}) was a popular open-source MoE inference framework.
It partitioned each layer's experts between the CPU and GPU, executed them in parallel, 
and assigned experts based on activation frequency to improve load balance.
For fair comparison, we used KTransformers v0.6.2 with the AVX2-only CPU backend and disabled expert deferral to ensure lossless inference.
(4) HybriMoE~\cite{hybrimoe} (\textbf{Hybri}) used prediction-guided dynamic partitioning.
It modeled CPU execution time as proportional to the number of input tokens, 
while treating GPU computation and PCIe transfer time as constant per expert, and used these models for load balancing.
In our experiments, HybriMoE used the same cache management 
and prefetching strategies as {\sys} to ensure a fair comparison of scheduling strategies.

Since FineMoE and HybriMoE were built on Python-based codebases with substantial interpreter overhead, 
we reproduced their methods on top of llama.cpp for fair performance comparison 
and carefully tuned them for optimal performance.
On the same MoE models, our reproductions achieved substantially higher performance than the original implementations.
For CPU/GPU hybrid inference systems, we used 8 threads for the CPU executor.

\stitle{Metrics.}
We evaluated prefill and decode performance separately,
following standard practice~\cite{ktransformers, yu2026taming, hybrimoe, MoE-APEX}.
We measured prefill throughput using \textbf{PPS} (prompts per second),
which was more intuitive than TTFT (time to first token),
and decode throughput using \textbf{TPS} (tokens per second).
Higher values indicated better performance for both metrics.
We assessed CPU-GPU load balance using \textbf{LBR} (load-balancing ratio),
the fraction of total hybrid expert-execution time during which
the CPU and GPU executed concurrently.
LBR ranged from 0 to 1, with higher values indicating better balance.
We measured the \textbf{Throughput} of co-executing workloads
in builds/min for compilation and images/s for ResNet-50 inference.
When reporting CPU/GPU expert fractions, 
we attributed each expert to the executor whose result was used in the output merge.

\subsection{Overall Performance}

We evaluated {\sys} and its baselines under Solo and the CPU co-execution
scenarios from Low to LP in Table~\ref{tab:workload-config}
to cover different resource-sharing conditions.
We used QW-3 and DS-2 with batch size 4 and QW-1 with batch size 1
to vary the number of expert tasks and their activation patterns.
Figure~\ref{fig:eval:overall} presented the results for all three configurations.
GPU contention was evaluated separately in \S\ref{sec:eval:gpu}.

\stitle{Solo.}
Compared with FineMoE, a GPU-only method, {\sys} achieved average speedups 
of 1.25$\times$ and 1.95$\times$ in prefill and decode, respectively, 
by executing experts on the CPU in parallel with GPU execution.
Compared with llama.cpp, the corresponding speedups were 2.65$\times$ and 1.62$\times$.
{\sys} reduced expert transfers through CPU execution in prefill and avoided sequential CPU and GPU execution in decode.

Compared with KTransformers, {\sys} achieved average speedups of 6.94$\times$ and 1.30$\times$ in prefill and decode, respectively.
KTransformers' static partitioning yielded average LBRs of 0.01 and 0.10 in prefill and decode, respectively,
compared with 0.93 and 0.97 for {\sys}.
Compared with HybriMoE, {\sys} achieved average speedups of 1.01$\times$ and 1.22$\times$, respectively.
Inaccuracies in HybriMoE's performance model limited its average LBRs to 0.85 and 0.39.
These results highlighted the benefit of dynamic scheduling even without co-executing programs.

\stitle{Co-execution with CPU workloads.}
In the Low, Medium, and High scenarios, we increased compilation concurrency 
while keeping the shared CPU set unchanged, 
simulating different levels of CPU contention for co-executing workloads.

Both HybriMoE and KTransformers failed to adapt to contention-induced CPU slowdowns 
because of inaccurate performance predictions and static expert partitioning, respectively.
As contention intensified, HybriMoE's average prefill LBR fell from 0.80 to 0.48.
Its decode LBR decreased for QW-3 and DS-2 but increased for QW-1.
For QW-1, HybriMoE assigned too many experts to the GPU in Solo;
CPU contention prolonged CPU execution, bringing the two paths closer in duration.
Overall, its decode LBR averaged only 0.35 across the three configurations and contention levels.
KTransformers' greater reliance on CPU execution and busy-waiting
synchronization further increased its sensitivity to contention.
In contrast, {\sys} achieved average prefill and decode LBRs of 0.89 and 0.79, respectively,
showing its adaptability to varying contention levels.
Across all three configurations and contention levels, 
{\sys} achieved average prefill and decode speedups of 1.44$\times$ and 1.62$\times$ over HybriMoE, 
and 8.91$\times$ and 3.18$\times$ over KTransformers.

\begin{figure}[t]
\centering
\begin{minipage}{\linewidth}
\centering
\includegraphics[width=1\linewidth]{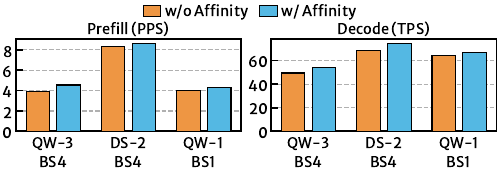}
\end{minipage} \\[2pt]
\begin{minipage}{1\linewidth}
\caption{\small{\emph{
Comparison of {\sys}'s prefill and decode performance with and without affinity-aware scheduling.
}}}
\label{fig:eval:breakdown-affinity-performance}
\end{minipage}
\end{figure}

Although FineMoE's throughput remained relatively stable because its expert execution did not compete with compilation for CPU time,
{\sys} achieved an average decode speedup of 1.67$\times$ over FineMoE.
In prefill, {\sys} achieved average speedups of 1.25$\times$ and 1.13$\times$ under Low and Medium contention, respectively,
but averaged only 81.2\% of FineMoE's performance under High contention.
FineMoE's cache policy was better suited to prefill, 
yielding an average cache hit rate 1.12$\times$ that of {\sys} under High contention.
Stronger contention reduced the contribution of CPU execution, making this cache advantage more consequential.
Llama.cpp also used a GPU-only method for prefill,
and {\sys} consistently outperformed it with an average prefill speedup of 2.74$\times$.
In decode, {\sys} achieved an average speedup of 6.34$\times$
over llama.cpp across the contention levels.

The Dynamic scenario showed similar inference performance trends.
{\sys} maintained high inference throughput and average prefill and decode
LBRs of 0.89 and 0.78, respectively,
demonstrating its adaptability to time-varying CPU contention.

Across these four co-execution scenarios, compilation achieved 98.6\% of its throughput 
under the GPU-only FineMoE on average when co-executed with {\sys}, 
whose blocking CPU synchronization allowed waiting workers to yield CPU time.
Compilation throughput was similar under {\sys} and HybriMoE,
which used the same CPU operators and exhibited similar CPU execution patterns.
Under {\sys}, compilation achieved 1.14$\times$ and 1.53$\times$
the throughput obtained under llama.cpp and KTransformers, respectively, on average.
Llama.cpp's CPU execution during decode and KTransformers' heavier CPU
workload and busy-waiting synchronization left fewer CPU cycles for compilation.

\stitle{Scheduling priorities.}
We used HP and LP to examine how {\sys} adapted to OS scheduling decisions,
with compilation assigned higher and lower priority than inference, respectively.
Both scenarios used the same compilation concurrency as Medium.

Under HP, compilation throughput was similar across inference systems
because the OS prioritized compilation.
Despite inference's lower priority, {\sys} maintained average prefill
and decode LBRs of 0.80 and 0.76, respectively,
and achieved corresponding average speedups of 17.58$\times$
and 7.62$\times$ over the baselines.
Under LP, inference performance followed a pattern similar to Solo,
as the OS prioritized inference over compilation.
Compared with {\sys}, FineMoE and llama.cpp achieved higher compilation
throughput but lower performance for the higher-priority inference workload.

\stitle{Summary.}
Across the three configurations and four baselines, {\sys} achieved
average prefill and decode speedups of 2.96$\times$ and
1.52$\times$ under Solo, respectively.
Across the six CPU co-execution scenarios, the corresponding average
speedups were 5.97$\times$ and 3.45$\times$,
while compilation achieved 1.08$\times$ the throughput obtained with the baselines on average.
These results demonstrated that {\sys} improved inference performance
while accommodating co-executing CPU workloads.

\subsection{Affinity-aware Scheduling}

\begin{figure}[t]
\centering
\begin{minipage}{\linewidth}
\centering
\includegraphics[width=1\linewidth]{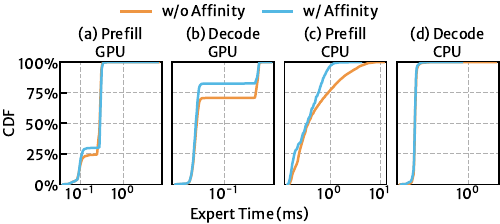}
\end{minipage} \\[2pt]
\begin{minipage}{1\linewidth}
\caption{\small{\emph{
The CDF of per-expert execution time on the GPU and CPU for {\sys} 
with and without affinity-aware scheduling during the prefill and decode 
stages when serving QW-3 with batch size 4.
}}}
\label{fig:eval:breakdown-affinity-cdf}
\end{minipage}
\end{figure}

Figure~\ref{fig:eval:breakdown-affinity-performance} showed that
affinity-aware scheduling achieved average speedups of
1.09$\times$ in prefill and 1.08$\times$ in decode.
To explain these gains, Figure~\ref{fig:eval:breakdown-affinity-cdf} compared the per-expert
execution-time distributions on the GPU and CPU with and without affinity-aware scheduling 
when serving QW-3 with batch size 4.

On the GPU, expert execution time was mainly determined by whether the expert parameters were cached.
This produced a two-stage CDF: the left stage corresponded to cache-hit experts, 
while the right stage corresponded to cache-miss experts.
Affinity-aware scheduling prioritized cache-hit experts for
GPU execution and gave frequently activated cache-miss experts
higher GPU priority to improve cache hit rates in subsequent iterations.
Together, these priorities increased the fraction of cache-hit
experts among GPU executions
(Figure~\ref{fig:eval:breakdown-affinity-cdf}(a) and (b)).

On the CPU, expert execution time increased with the number of input tokens, 
so affinity-aware scheduling prioritized experts with fewer input tokens for CPU execution.
In prefill, each forward pass processed many tokens, producing
substantial variation in token counts and execution times across experts.
The scheduling policy therefore reduced the number of long-running CPU executions
(Figure~\ref{fig:eval:breakdown-affinity-cdf}(c)).
In decode, each forward pass processed fewer tokens, so expert execution times were more similar.
Nevertheless, affinity-aware scheduling still reduced the tail execution times on the CPU (Figure~\ref{fig:eval:breakdown-affinity-cdf}(d)).

Overall, affinity-aware scheduling improved the performance of {\sys} 
by prioritizing experts with better affinity to the target processor, either GPU or CPU,
which shifted the per-expert execution-time distribution to the left.

\subsection{Cancellable Tail Racing}

\begin{figure}[t]
\centering
\begin{minipage}{\linewidth}
\centering
\includegraphics[width=1\linewidth]{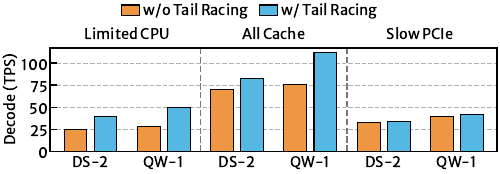}
\end{minipage} \\[2pt]
\begin{minipage}{1\linewidth}
\caption{\small{\emph{
Comparison of performance for {\sys} 
with and without cancellable tail racing in three straggler scenarios
when serving the DS-2 and QW-1 models with batch size 1.
}}}
\label{fig:eval:breakdown-backup}
\end{minipage}
\end{figure}

We evaluated cancellable tail racing on DS-2 and QW-1 with batch size 1 under three controlled straggler scenarios.
We focused on decode because fewer expert tasks made this stage more sensitive to stragglers.
(1) \textbf{Limited CPU:}
We used the High contention scenario in Table~\ref{tab:workload-config},
where CPU contention made the CPU executor the straggler.
(2) \textbf{All Cache:}
We remapped activated experts to GPU-cached experts to achieve a 100\% cache hit rate.
The faster GPU execution left the CPU executor as the straggler.
(3) \textbf{Slow PCIe:}
We increased each GPU transfer to four times its original size
to emulate a slow or contended PCIe link, making the GPU executor the straggler.

Figure~\ref{fig:eval:breakdown-backup} compared decode throughput with and without tail racing.
In Limited CPU and All Cache, tail racing allowed the GPU to complete straggling CPU tasks.
It reduced the average CPU expert fraction from 22.43\% to 3.49\%
and achieved average speedups of 1.68$\times$ and 1.32$\times$, respectively.
In Slow PCIe, tail racing allowed the CPU to complete straggling GPU tasks,
increasing the average CPU expert fraction from 39.69\% to 40.18\%.
However, the average GPU cache hit rate fell from 72.15\% to 64.91\% in this scenario.
The additional parameter transfers offset part of the benefit of mitigating stragglers,
but tail racing still achieved an average decode speedup of 1.05$\times$
over the version without tail racing.

We also evaluated tail racing under Solo
using QW-3 and DS-2 with batch size 4 and QW-1 with batch size 1.
Across these configurations, it achieved average speedups
of 1.12$\times$ in prefill and 1.01$\times$ in decode.

\begin{figure}[t]
\centering
\begin{minipage}{\linewidth}
\centering
\includegraphics[width=1\linewidth]{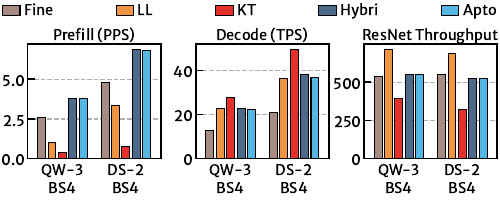}
\end{minipage} \\[2pt]
\begin{minipage}{1\linewidth}
\caption{\small{\emph{
Comparison of the inference performance and ResNet-50 throughput for {\sys} and its baselines 
when serving the DS-2 and QW-3 models with batch size 4 under the GPU scenario.
}}}
\label{fig:eval:overall-gpu}
\end{minipage}
\end{figure}

\subsection{GPU Contention}
\label{sec:eval:gpu}

Although GPU contention was less common on PCs, we used a controlled experiment to verify 
that {\sys} adapted symmetrically when GPU capacity varied.
We evaluated {\sys} and its baselines on DS-2 and QW-3 with batch size 4
under the GPU scenario in Table~\ref{tab:workload-config}.
Compared with Solo, {\sys} reduced the GPU expert fraction from 81.84\% to 80.50\% in prefill and
from 79.60\% to 68.15\% in decode in response to GPU contention.
It maintained LBRs above 0.83 and 0.90, respectively.

As shown in Figure~\ref{fig:eval:overall-gpu},
compared with FineMoE, {\sys} achieved average speedups of
1.44$\times$ and 1.75$\times$ in prefill and decode,
respectively, with similar ResNet-50 throughput.
Compared with llama.cpp, {\sys} achieved an average prefill speedup
of 2.89$\times$ and similar decode throughput,
while retaining 76.79\% of the ResNet-50 throughput obtained with llama.cpp.
Llama.cpp's sequential execution left more GPU time for ResNet-50, 
while {\sys}'s parallel CPU/GPU execution increased competition for GPU resources.

Compared with KTransformers, {\sys} achieved 9.23$\times$ the prefill throughput and
77.06\% of the decode throughput, while delivering 1.51$\times$ the ResNet-50 throughput.
KTransformers retained more of its Solo decode throughput, but interfered more with ResNet-50.
Its use of CUDA graphs contributed to this trade-off
by reducing gaps between GPU operations and leaving fewer opportunities for ResNet-50 execution.
In decode, HybriMoE's inaccurate performance model assigned too many experts to the CPU for both models, 
leaving the GPU underutilized in Solo.
GPU contention offset this imbalance by slowing GPU execution. 
HybriMoE achieved average LBRs of 0.85 and 0.97 in prefill and decode, respectively, close to those of {\sys}.
Their similar expert partitioning and CPU/GPU load balance resulted in comparable inference and ResNet-50 throughput.

\subsection{Case Study: Multi-agent Application}

\begin{figure}[t]
\centering
\begin{minipage}{\linewidth}
\centering
\includegraphics[width=1\linewidth]{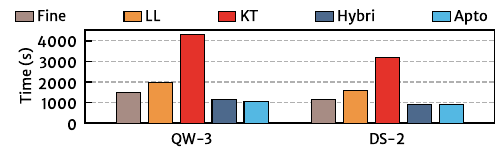}
\end{minipage} \\[2pt]
\begin{minipage}{1\linewidth}
\caption{\small{\emph{
Completion time of the multi-agent application 
using QW-3 and DS-2 models served by {\sys} and its baselines.
The number of available CPU cores was set to 12.
}}}
\label{fig:eval:sweagent}
\end{minipage}
\end{figure}

To evaluate end-to-end performance in a multi-agent application,
we replayed the traces discussed in Section~\ref{sec:bg} with four concurrent agent workers.
We deployed {\sys} and each baseline in turn as a local LLM inference server
and routed the recorded LLM requests to it.
For each request, we fixed the output length to match that of the corresponding response in the original trace.
After inference completed, each worker continued with the recorded
agent actions rather than using the locally generated response to determine subsequent actions.
Figure~\ref{fig:eval:sweagent} showed the completion times for QW-3 and DS-2.
Across both models and all baselines, {\sys} consistently achieved
the shortest completion time, with speedups of
1.04$\times$--4.07$\times$ and an average of 2.02$\times$.

\subsection{UMA-based PCs}
\label{sec:eval:uma}

\begin{figure}[t]
\centering
\begin{minipage}{\linewidth}
\centering
\includegraphics[width=1\linewidth]{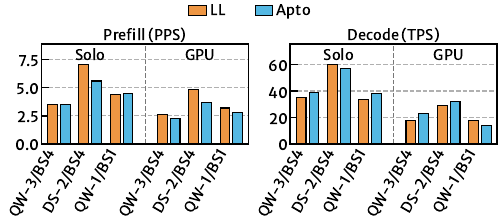}
\end{minipage} \\[2pt]
\begin{minipage}{1\linewidth}
\caption{\small{\emph{
Comparison of prefill and decode performance for {\sys} and llama.cpp on a UMA-based PC 
under Solo and GPU scenarios.
}}}
\label{fig:eval:uma}
\end{minipage}
\end{figure}

We evaluated {\sys}'s performance and adaptability on NVIDIA DGX Spark.
Since all expert parameters fit in shared memory without offloading,
we used GPU-only llama.cpp as the baseline.
We evaluated QW-3 and DS-2 with batch size 4 and QW-1 with batch size 1
under the Solo and GPU scenarios in Table~\ref{tab:workload-config}.
Figure~\ref{fig:eval:uma} showed their prefill and decode throughput.

Under Solo, {\sys} maintained a near-zero CPU expert fraction
and achieved an average decode speedup of 1.07$\times$ over llama.cpp.
Under GPU contention, it shifted work to the CPU when GPU progress stalled.
For QW-3 and DS-2, GPU expert fractions dropped to 98.93\% and 95.53\%,
with decode speedups of 1.27$\times$ and 1.09$\times$, respectively.
For QW-1, limited scheduling flexibility left decode performance at 81.03\% of llama.cpp's.
ResNet-50 throughput alongside {\sys} averaged 93.93\% of that alongside llama.cpp
across the three configurations because of shared memory bandwidth contention.

In prefill, {\sys} achieved an average of 88.37\% of llama.cpp's performance
across the three configurations and both scenarios.
This lower performance resulted from the use of pinned host memory
to make expert parameters accessible to both the CPU and GPU.
The coherence requirements of these shared accesses introduced overhead compared with device memory.

\section{Related Work}
\label{sec:related}

\nostitle{Serving MoE models under limited GPU memory.}
GPU-only methods~\cite{moe-infinity, mixtral_offloading, yu2026taming, song2024promoe}
reduce offloading overhead through expert caching and prefetching.
{\sys} complements these techniques with CPU execution to further improve performance.
Prior CPU/GPU hybrid methods~\cite{hybrimoe, LayerScope, ktransformers, kamahori2024fiddler, cao2025moe}
are self-centered, optimizing resource utilization under exclusive resource access.
In contrast, {\sys} enables self-adaptive hybrid execution through dynamic scheduling,
adapting to runtime resource contention.

\stitle{Algorithm-system co-design for MoE inference.}
Prior work~\cite{pregatemoe, kong2024swapmoe, MoE-APEX, ktransformers, zhu2026smoealgorithmsystemcodesignpushing}
modifies expert selection, numerical precision, or execution dependencies
to reduce data transfer or improve computation overlap.
Such changes alter the original model computation and unavoidably affect model accuracy.
In contrast, {\sys} optimizes expert execution through dynamic scheduling
while preserving the original inference computation flow.

\stitle{General-purpose dynamic scheduling.}
Dynamic scheduling is widely used to balance workloads and improve resource utilization
in distributed clusters~\cite{dean2004mapreduce, zaharia2008improving},
multicore systems~\cite{blumofe1999scheduling, chen2010tmr},
and heterogeneous platforms~\cite{augonnet2011starpu, kaleem2014adaptive}.
{\sys} specializes dynamic scheduling for MoE CPU/GPU hybrid inference on shared PCs,
accounting for expert characteristics while adapting the CPU/GPU work split
to runtime resource contention. 
\section{Conclusion}

This paper presents {\sys}, a self-adaptive CPU/GPU hybrid inference system for MoE models on shared PCs.
Through dynamic scheduling, {\sys} adapts expert assignments to execution progress and runtime resource contention.
Our evaluation demonstrates its efficiency and adaptability under diverse colocated workloads.

\small{
\bibliographystyle{ACM-Reference-Format} \bibliography{main}


\begin{thebibliography}{66}


\ifx \showCODEN    \undefined \def \showCODEN     #1{\unskip}     \fi
\ifx \showISBNx    \undefined \def \showISBNx     #1{\unskip}     \fi
\ifx \showISBNxiii \undefined \def \showISBNxiii  #1{\unskip}     \fi
\ifx \showISSN     \undefined \def \showISSN      #1{\unskip}     \fi
\ifx \showLCCN     \undefined \def \showLCCN      #1{\unskip}     \fi
\ifx \shownote     \undefined \def \shownote      #1{#1}          \fi
\ifx \showarticletitle \undefined \def \showarticletitle #1{#1}   \fi
\ifx \showURL      \undefined \def \showURL       {\relax}        \fi
\providecommand\bibfield[2]{#2}
\providecommand\bibinfo[2]{#2}
\providecommand\natexlab[1]{#1}
\providecommand\showeprint[2][]{arXiv:#2}

\bibitem[sha(2024)]%
        {sharegpt}
 \bibinfo{year}{2024}\natexlab{}.
\newblock \bibinfo{title}{anon8231489123/{ShareGPT}\_Vicuna\_unfiltered ·
  Datasets at Hugging Face}.
\newblock


\bibitem[qw1(2024)]%
        {qw1}
 \bibinfo{year}{2024}\natexlab{}.
\newblock \bibinfo{booktitle}{\emph{Qwen/Qwen1.5-{MoE}-A2.7B-Chat · Hugging
  Face}}.
\newblock
\urldef\tempurl%
\url{https://huggingface.co/Qwen/Qwen1.5-MoE-A2.7B-Chat}
\showURL{%
\tempurl}


\bibitem[zst(2026)]%
        {zstd}
 \bibinfo{year}{2026}\natexlab{}.
\newblock \bibinfo{title}{facebook/zstd: Zstandard - Fast real-time compression
  algorithm}.
\newblock \bibinfo{howpublished}{\url{https://github.com/facebook/zstd}}.
\newblock


\bibitem[cod(2026)]%
        {codex}
 \bibinfo{year}{2026}\natexlab{}.
\newblock \bibinfo{title}{openai/codex: Lightweight coding agent that runs in
  your terminal}.
\newblock \bibinfo{howpublished}{\url{https://github.com/openai/codex}}.
\newblock


\bibitem[{\'A}lvarez et~al\mbox{.}(2021)]%
        {alvarez2021advanced}
\bibfield{author}{\bibinfo{person}{David {\'A}lvarez}, \bibinfo{person}{Kevin
  Sala}, \bibinfo{person}{Marcos Maro{\~n}as}, \bibinfo{person}{Aleix Roca},
  {and} \bibinfo{person}{Vincen{\c{c}} Beltran}.}
  \bibinfo{year}{2021}\natexlab{}.
\newblock \showarticletitle{Advanced synchronization techniques for task-based
  runtime systems}. In \bibinfo{booktitle}{\emph{Proceedings of the 26th ACM
  SIGPLAN Symposium on Principles and Practice of Parallel Programming}}.
  \bibinfo{pages}{334--347}.
\newblock


\bibitem[Ananthanarayanan et~al\mbox{.}(2013)]%
        {ananthanarayanan2013effective}
\bibfield{author}{\bibinfo{person}{Ganesh Ananthanarayanan},
  \bibinfo{person}{Ali Ghodsi}, \bibinfo{person}{Scott Shenker}, {and}
  \bibinfo{person}{Ion Stoica}.} \bibinfo{year}{2013}\natexlab{}.
\newblock \showarticletitle{Effective straggler mitigation: Attack of the
  clones}. In \bibinfo{booktitle}{\emph{10th USENIX Symposium on Networked
  Systems Design and Implementation (NSDI 13)}}. \bibinfo{pages}{185--198}.
\newblock


\bibitem[{Apple Inc}(2025a)]%
        {appleincMacStudio2025}
\bibfield{author}{\bibinfo{person}{{Apple Inc}}.}
  \bibinfo{year}{2025}\natexlab{a}.
\newblock \bibinfo{title}{Mac {{Studio}}}.
\newblock \bibinfo{howpublished}{\url{https://support.apple.com/en-us/122211}}.
\newblock


\bibitem[{Apple Inc}(2025b)]%
        {appleincMacBookPro2025}
\bibfield{author}{\bibinfo{person}{{Apple Inc}}.}
  \bibinfo{year}{2025}\natexlab{b}.
\newblock \bibinfo{title}{{{MacBook Pro}}}.
\newblock \bibinfo{howpublished}{\url{https://support.apple.com/en-us/125405}}.
\newblock


\bibitem[Augonnet et~al\mbox{.}(2010)]%
        {augonnet2010data}
\bibfield{author}{\bibinfo{person}{C{\'e}dric Augonnet},
  \bibinfo{person}{J{\'e}r{\^o}me Clet-Ortega}, \bibinfo{person}{Samuel
  Thibault}, {and} \bibinfo{person}{Raymond Namyst}.}
  \bibinfo{year}{2010}\natexlab{}.
\newblock \showarticletitle{Data-aware task scheduling on multi-accelerator
  based platforms}. In \bibinfo{booktitle}{\emph{2010 IEEE 16th international
  conference on parallel and distributed systems}}. IEEE,
  \bibinfo{pages}{291--298}.
\newblock


\bibitem[Augonnet et~al\mbox{.}(2011)]%
        {augonnet2011starpu}
\bibfield{author}{\bibinfo{person}{C{\'e}dric Augonnet},
  \bibinfo{person}{Samuel Thibault}, \bibinfo{person}{Raymond Namyst}, {and}
  \bibinfo{person}{Pierre-Andr{\'e} Wacrenier}.}
  \bibinfo{year}{2011}\natexlab{}.
\newblock \showarticletitle{StarPU: a unified platform for task scheduling on
  heterogeneous multicore architectures}.
\newblock \bibinfo{journal}{\emph{Concurrency and computation: practice and
  experience}} \bibinfo{volume}{23}, \bibinfo{number}{2}
  (\bibinfo{year}{2011}), \bibinfo{pages}{187--198}.
\newblock


\bibitem[{Beelink}(2025)]%
        {beelinkBeelinkGTR9Pro2025}
\bibfield{author}{\bibinfo{person}{{Beelink}}.}
  \bibinfo{year}{2025}\natexlab{}.
\newblock \bibinfo{title}{Beelink {{GTR9 Pro AMD Ryzen AI Max}}+ 395
  {{Processor}}}.
\newblock
  \bibinfo{howpublished}{\url{https://www.bee-link.com/products/beelink-gtr9-pro-amd-ryzen-ai-max-395}}.
\newblock


\bibitem[Blumofe and Leiserson(1999)]%
        {blumofe1999scheduling}
\bibfield{author}{\bibinfo{person}{Robert~D Blumofe} {and}
  \bibinfo{person}{Charles~E Leiserson}.} \bibinfo{year}{1999}\natexlab{}.
\newblock \showarticletitle{Scheduling multithreaded computations by work
  stealing}.
\newblock \bibinfo{journal}{\emph{Journal of the ACM (JACM)}}
  \bibinfo{volume}{46}, \bibinfo{number}{5} (\bibinfo{year}{1999}),
  \bibinfo{pages}{720--748}.
\newblock


\bibitem[{\c{C}}al{\i}ky{\i}lmaz et~al\mbox{.}(2025)]%
        {ccalikyilmaz2025optima}
\bibfield{author}{\bibinfo{person}{Umut {\c{C}}al{\i}ky{\i}lmaz},
  \bibinfo{person}{Nitin Nayak}, \bibinfo{person}{Jinghua Groppe}, {and}
  \bibinfo{person}{Sven Groppe}.} \bibinfo{year}{2025}\natexlab{}.
\newblock \showarticletitle{OptiMA: a transaction-based framework with
  throughput optimization for very complex multi-agent systems}.
\newblock \bibinfo{journal}{\emph{arXiv preprint arXiv:2511.03761}}
  (\bibinfo{year}{2025}).
\newblock


\bibitem[Cao et~al\mbox{.}(2025)]%
        {cao2025moe}
\bibfield{author}{\bibinfo{person}{Shiyi Cao}, \bibinfo{person}{Shu Liu},
  \bibinfo{person}{Tyler Griggs}, \bibinfo{person}{Peter Schafhalter},
  \bibinfo{person}{Xiaoxuan Liu}, \bibinfo{person}{Ying Sheng},
  \bibinfo{person}{Joseph~E Gonzalez}, \bibinfo{person}{Matei Zaharia}, {and}
  \bibinfo{person}{Ion Stoica}.} \bibinfo{year}{2025}\natexlab{}.
\newblock \showarticletitle{Moe-lightning: High-throughput moe inference on
  memory-constrained gpus}. In \bibinfo{booktitle}{\emph{Proceedings of the
  30th ACM International Conference on Architectural Support for Programming
  Languages and Operating Systems, Volume 1}}. \bibinfo{pages}{715--730}.
\newblock


\bibitem[Chase and Lev(2005)]%
        {chase2005dynamic}
\bibfield{author}{\bibinfo{person}{David Chase} {and} \bibinfo{person}{Yossi
  Lev}.} \bibinfo{year}{2005}\natexlab{}.
\newblock \showarticletitle{Dynamic circular work-stealing deque}. In
  \bibinfo{booktitle}{\emph{Proceedings of the seventeenth annual ACM symposium
  on Parallelism in algorithms and architectures}}. \bibinfo{pages}{21--28}.
\newblock


\bibitem[Chen et~al\mbox{.}(2025b)]%
        {ktransformers}
\bibfield{author}{\bibinfo{person}{Hongtao Chen}, \bibinfo{person}{Weiyu Xie},
  \bibinfo{person}{Boxin Zhang}, \bibinfo{person}{Jingqi Tang},
  \bibinfo{person}{Jiahao Wang}, \bibinfo{person}{Jianwei Dong},
  \bibinfo{person}{Shaoyuan Chen}, \bibinfo{person}{Ziwei Yuan},
  \bibinfo{person}{Chen Lin}, \bibinfo{person}{Chengyu Qiu}, {et~al\mbox{.}}}
  \bibinfo{year}{2025}\natexlab{b}.
\newblock \showarticletitle{Ktransformers: Unleashing the full potential of
  cpu/gpu hybrid inference for moe models}. In
  \bibinfo{booktitle}{\emph{Proceedings of the ACM SIGOPS 31st Symposium on
  Operating Systems Principles}}. \bibinfo{pages}{1014--1029}.
\newblock


\bibitem[Chen et~al\mbox{.}(2025a)]%
        {chen2025characterizing}
\bibfield{author}{\bibinfo{person}{Le Chen}, \bibinfo{person}{Dahu Feng},
  \bibinfo{person}{Erhu Feng}, \bibinfo{person}{Yingrui Wang},
  \bibinfo{person}{Rong Zhao}, \bibinfo{person}{Yubin Xia},
  \bibinfo{person}{Pinjie Xu}, {and} \bibinfo{person}{Haibo Chen}.}
  \bibinfo{year}{2025}\natexlab{a}.
\newblock \showarticletitle{Characterizing mobile soc for accelerating
  heterogeneous llm inference}. In \bibinfo{booktitle}{\emph{Proceedings of the
  ACM SIGOPS 31st Symposium on Operating Systems Principles}}.
  \bibinfo{pages}{359--374}.
\newblock


\bibitem[Chen et~al\mbox{.}(2010)]%
        {chen2010tmr}
\bibfield{author}{\bibinfo{person}{Rong Chen}, \bibinfo{person}{Haibo Chen},
  {and} \bibinfo{person}{Binyu Zang}.} \bibinfo{year}{2010}\natexlab{}.
\newblock \showarticletitle{Tiled-MapReduce: optimizing resource usages of
  data-parallel applications on multicore with tiling}. In
  \bibinfo{booktitle}{\emph{Proceedings of the 19th International Conference on
  Parallel Architectures and Compilation Techniques}} (Vienna, Austria)
  \emph{(\bibinfo{series}{PACT '10})}. \bibinfo{publisher}{Association for
  Computing Machinery}, \bibinfo{address}{New York, NY, USA},
  \bibinfo{pages}{523–534}.
\newblock
\showISBNx{9781450301787}
\href{https://doi.org/10.1145/1854273.1854337}{doi:\nolinkurl{10.1145/1854273.1854337}}


\bibitem[Corporation(2026)]%
        {cudaProgrammingGuide}
\bibfield{author}{\bibinfo{person}{NVIDIA Corporation}.}
  \bibinfo{year}{2026}\natexlab{}.
\newblock \bibinfo{title}{CUDA Programming Guide}.
\newblock
  \bibinfo{howpublished}{\url{https://docs.nvidia.com/cuda/cuda-programming-guide}}.
\newblock


\bibitem[Dai et~al\mbox{.}(2024)]%
        {dai_deepseekmoe_2024}
\bibfield{author}{\bibinfo{person}{Damai Dai}, \bibinfo{person}{Chengqi Deng},
  \bibinfo{person}{Chenggang Zhao}, \bibinfo{person}{R.~X. Xu},
  \bibinfo{person}{Huazuo Gao}, \bibinfo{person}{Deli Chen},
  \bibinfo{person}{Jiashi Li}, \bibinfo{person}{Wangding Zeng},
  \bibinfo{person}{Xingkai Yu}, \bibinfo{person}{Y. Wu},
  \bibinfo{person}{Zhenda Xie}, \bibinfo{person}{Y.~K. Li},
  \bibinfo{person}{Panpan Huang}, \bibinfo{person}{Fuli Luo},
  \bibinfo{person}{Chong Ruan}, \bibinfo{person}{Zhifang Sui}, {and}
  \bibinfo{person}{Wenfeng Liang}.} \bibinfo{year}{2024}\natexlab{}.
\newblock \bibinfo{title}{{DeepSeekMoE}: Towards Ultimate Expert Specialization
  in Mixture-of-Experts Language Models}.
\newblock
\showeprint[arxiv]{2401.06066}
\href{https://doi.org/10.48550/arXiv.2401.06066}{doi:\nolinkurl{10.48550/arXiv.2401.06066}}


\bibitem[Dean et~al\mbox{.}(2004)]%
        {dean2004mapreduce}
\bibfield{author}{\bibinfo{person}{Jeffrey Dean}, \bibinfo{person}{Sanjay
  Ghemawat}, {et~al\mbox{.}}} \bibinfo{year}{2004}\natexlab{}.
\newblock \showarticletitle{Mapreduce: simplified data processing on large
  clusters.}. In \bibinfo{booktitle}{\emph{osdi}}, Vol.~\bibinfo{volume}{4}.
  USA, \bibinfo{pages}{137--150}.
\newblock


\bibitem[{DeepSeek}-{AI} et~al\mbox{.}(2024)]%
        {deepseek-ai_deepseek-v2_2024}
\bibfield{author}{\bibinfo{person}{{DeepSeek}-{AI}}, \bibinfo{person}{Aixin
  Liu}, \bibinfo{person}{Bei Feng}, \bibinfo{person}{Bin Wang},
  \bibinfo{person}{Bingxuan Wang}, \bibinfo{person}{Bo Liu},
  \bibinfo{person}{Chenggang Zhao}, \bibinfo{person}{Chengqi Dengr},
  \bibinfo{person}{Chong Ruan}, \bibinfo{person}{Damai Dai},
  \bibinfo{person}{Daya Guo}, \bibinfo{person}{Dejian Yang},
  \bibinfo{person}{Deli Chen}, \bibinfo{person}{Dongjie Ji},
  \bibinfo{person}{Erhang Li}, \bibinfo{person}{Fangyun Lin},
  \bibinfo{person}{Fuli Luo}, \bibinfo{person}{Guangbo Hao},
  \bibinfo{person}{Guanting Chen}, \bibinfo{person}{Guowei Li},
  \bibinfo{person}{H. Zhang}, \bibinfo{person}{Hanwei Xu}, \bibinfo{person}{Hao
  Yang}, \bibinfo{person}{Haowei Zhang}, \bibinfo{person}{Honghui Ding},
  \bibinfo{person}{Huajian Xin}, \bibinfo{person}{Huazuo Gao},
  \bibinfo{person}{Hui Li}, \bibinfo{person}{Hui Qu}, \bibinfo{person}{J.~L.
  Cai}, \bibinfo{person}{Jian Liang}, \bibinfo{person}{Jianzhong Guo},
  \bibinfo{person}{Jiaqi Ni}, \bibinfo{person}{Jiashi Li}, \bibinfo{person}{Jin
  Chen}, \bibinfo{person}{Jingyang Yuan}, \bibinfo{person}{Junjie Qiu},
  \bibinfo{person}{Junxiao Song}, \bibinfo{person}{Kai Dong},
  \bibinfo{person}{Kaige Gao}, \bibinfo{person}{Kang Guan},
  \bibinfo{person}{Lean Wang}, \bibinfo{person}{Lecong Zhang},
  \bibinfo{person}{Lei Xu}, \bibinfo{person}{Leyi Xia}, \bibinfo{person}{Liang
  Zhao}, \bibinfo{person}{Liyue Zhang}, \bibinfo{person}{Meng Li},
  \bibinfo{person}{Miaojun Wang}, \bibinfo{person}{Mingchuan Zhang},
  \bibinfo{person}{Minghua Zhang}, \bibinfo{person}{Minghui Tang},
  \bibinfo{person}{Mingming Li}, \bibinfo{person}{Ning Tian},
  \bibinfo{person}{Panpan Huang}, \bibinfo{person}{Peiyi Wang},
  \bibinfo{person}{Peng Zhang}, \bibinfo{person}{Qihao Zhu},
  \bibinfo{person}{Qinyu Chen}, \bibinfo{person}{Qiushi Du},
  \bibinfo{person}{R.~J. Chen}, \bibinfo{person}{R.~L. Jin},
  \bibinfo{person}{Ruiqi Ge}, \bibinfo{person}{Ruizhe Pan},
  \bibinfo{person}{Runxin Xu}, \bibinfo{person}{Ruyi Chen},
  \bibinfo{person}{S.~S. Li}, \bibinfo{person}{Shanghao Lu},
  \bibinfo{person}{Shangyan Zhou}, \bibinfo{person}{Shanhuang Chen},
  \bibinfo{person}{Shaoqing Wu}, \bibinfo{person}{Shengfeng Ye},
  \bibinfo{person}{Shirong Ma}, \bibinfo{person}{Shiyu Wang},
  \bibinfo{person}{Shuang Zhou}, \bibinfo{person}{Shuiping Yu},
  \bibinfo{person}{Shunfeng Zhou}, \bibinfo{person}{Size Zheng},
  \bibinfo{person}{T. Wang}, \bibinfo{person}{Tian Pei}, \bibinfo{person}{Tian
  Yuan}, \bibinfo{person}{Tianyu Sun}, \bibinfo{person}{W.~L. Xiao},
  \bibinfo{person}{Wangding Zeng}, \bibinfo{person}{Wei An},
  \bibinfo{person}{Wen Liu}, \bibinfo{person}{Wenfeng Liang},
  \bibinfo{person}{Wenjun Gao}, \bibinfo{person}{Wentao Zhang},
  \bibinfo{person}{X.~Q. Li}, \bibinfo{person}{Xiangyue Jin},
  \bibinfo{person}{Xianzu Wang}, \bibinfo{person}{Xiao Bi},
  \bibinfo{person}{Xiaodong Liu}, \bibinfo{person}{Xiaohan Wang},
  \bibinfo{person}{Xiaojin Shen}, \bibinfo{person}{Xiaokang Chen},
  \bibinfo{person}{Xiaosha Chen}, \bibinfo{person}{Xiaotao Nie},
  \bibinfo{person}{Xiaowen Sun}, \bibinfo{person}{Xiaoxiang Wang},
  \bibinfo{person}{Xin Liu}, \bibinfo{person}{Xin Xie},
  \bibinfo{person}{Xingkai Yu}, \bibinfo{person}{Xinnan Song},
  \bibinfo{person}{Xinyi Zhou}, \bibinfo{person}{Xinyu Yang},
  \bibinfo{person}{Xuan Lu}, \bibinfo{person}{Xuecheng Su}, \bibinfo{person}{Y.
  Wu}, \bibinfo{person}{Y.~K. Li}, \bibinfo{person}{Y.~X. Wei},
  \bibinfo{person}{Y.~X. Zhu}, \bibinfo{person}{Yanhong Xu},
  \bibinfo{person}{Yanping Huang}, \bibinfo{person}{Yao Li},
  \bibinfo{person}{Yao Zhao}, \bibinfo{person}{Yaofeng Sun},
  \bibinfo{person}{Yaohui Li}, \bibinfo{person}{Yaohui Wang},
  \bibinfo{person}{Yi Zheng}, \bibinfo{person}{Yichao Zhang},
  \bibinfo{person}{Yiliang Xiong}, \bibinfo{person}{Yilong Zhao},
  \bibinfo{person}{Ying He}, \bibinfo{person}{Ying Tang},
  \bibinfo{person}{Yishi Piao}, \bibinfo{person}{Yixin Dong},
  \bibinfo{person}{Yixuan Tan}, \bibinfo{person}{Yiyuan Liu},
  \bibinfo{person}{Yongji Wang}, \bibinfo{person}{Yongqiang Guo},
  \bibinfo{person}{Yuchen Zhu}, \bibinfo{person}{Yuduan Wang},
  \bibinfo{person}{Yuheng Zou}, \bibinfo{person}{Yukun Zha},
  \bibinfo{person}{Yunxian Ma}, \bibinfo{person}{Yuting Yan},
  \bibinfo{person}{Yuxiang You}, \bibinfo{person}{Yuxuan Liu},
  \bibinfo{person}{Z.~Z. Ren}, \bibinfo{person}{Zehui Ren},
  \bibinfo{person}{Zhangli Sha}, \bibinfo{person}{Zhe Fu},
  \bibinfo{person}{Zhen Huang}, \bibinfo{person}{Zhen Zhang},
  \bibinfo{person}{Zhenda Xie}, \bibinfo{person}{Zhewen Hao},
  \bibinfo{person}{Zhihong Shao}, \bibinfo{person}{Zhiniu Wen},
  \bibinfo{person}{Zhipeng Xu}, \bibinfo{person}{Zhongyu Zhang},
  \bibinfo{person}{Zhuoshu Li}, \bibinfo{person}{Zihan Wang},
  \bibinfo{person}{Zihui Gu}, \bibinfo{person}{Zilin Li}, {and}
  \bibinfo{person}{Ziwei Xie}.} \bibinfo{year}{2024}\natexlab{}.
\newblock \bibinfo{title}{{DeepSeek}-V2: A Strong, Economical, and Efficient
  Mixture-of-Experts Language Model}.
\newblock
\showeprint[arxiv]{2405.04434}
\href{https://doi.org/10.48550/arXiv.2405.04434}{doi:\nolinkurl{10.48550/arXiv.2405.04434}}


\bibitem[Eliseev and Mazur(2023)]%
        {mixtral_offloading}
\bibfield{author}{\bibinfo{person}{Artyom Eliseev} {and} \bibinfo{person}{Denis
  Mazur}.} \bibinfo{year}{2023}\natexlab{}.
\newblock \bibinfo{title}{Fast Inference of Mixture-of-Experts Language Models
  with Offloading}.
\newblock
\showeprint[arxiv]{2312.17238}~[cs.LG]
\urldef\tempurl%
\url{https://arxiv.org/abs/2312.17238}
\showURL{%
\tempurl}


\bibitem[Fourney et~al\mbox{.}(2024)]%
        {fourney2024magentic}
\bibfield{author}{\bibinfo{person}{Adam Fourney}, \bibinfo{person}{Gagan
  Bansal}, \bibinfo{person}{Hussein Mozannar}, \bibinfo{person}{Cheng Tan},
  \bibinfo{person}{Eduardo Salinas}, \bibinfo{person}{Friederike Niedtner},
  \bibinfo{person}{Grace Proebsting}, \bibinfo{person}{Griffin Bassman},
  \bibinfo{person}{Jack Gerrits}, \bibinfo{person}{Jacob Alber},
  {et~al\mbox{.}}} \bibinfo{year}{2024}\natexlab{}.
\newblock \showarticletitle{Magentic-one: A generalist multi-agent system for
  solving complex tasks}.
\newblock \bibinfo{journal}{\emph{arXiv preprint arXiv:2411.04468}}
  (\bibinfo{year}{2024}).
\newblock


\bibitem[Gerganov(2024)]%
        {llamacpp}
\bibfield{author}{\bibinfo{person}{Georgi Gerganov}.}
  \bibinfo{year}{2024}\natexlab{}.
\newblock \bibinfo{title}{ggerganov/llama.cpp}.
\newblock
\urldef\tempurl%
\url{https://github.com/ggerganov/llama.cpp}
\showURL{%
\tempurl}
\newblock
\shownote{original-date: 2023-03-10T18:58:00Z}.


\bibitem[He et~al\mbox{.}(2016)]%
        {he2016deep}
\bibfield{author}{\bibinfo{person}{Kaiming He}, \bibinfo{person}{Xiangyu
  Zhang}, \bibinfo{person}{Shaoqing Ren}, {and} \bibinfo{person}{Jian Sun}.}
  \bibinfo{year}{2016}\natexlab{}.
\newblock \showarticletitle{Deep residual learning for image recognition}. In
  \bibinfo{booktitle}{\emph{Proceedings of the IEEE conference on computer
  vision and pattern recognition}}. \bibinfo{pages}{770--778}.
\newblock


\bibitem[Hwang et~al\mbox{.}(2024)]%
        {pregatemoe}
\bibfield{author}{\bibinfo{person}{Ranggi Hwang}, \bibinfo{person}{Jianyu Wei},
  \bibinfo{person}{Shijie Cao}, \bibinfo{person}{Changho Hwang},
  \bibinfo{person}{Xiaohu Tang}, \bibinfo{person}{Ting Cao}, {and}
  \bibinfo{person}{Mao Yang}.} \bibinfo{year}{2024}\natexlab{}.
\newblock \showarticletitle{Pre-gated MoE: An Algorithm-System Co-Design for
  Fast and Scalable Mixture-of-Expert Inference}. In
  \bibinfo{booktitle}{\emph{2024 ACM/IEEE 51st Annual International Symposium
  on Computer Architecture (ISCA)}}. \bibinfo{pages}{1018--1031}.
\newblock
\href{https://doi.org/10.1109/ISCA59077.2024.00078}{doi:\nolinkurl{10.1109/ISCA59077.2024.00078}}


\bibitem[{Intel Corporation}(2025)]%
        {intelcorporationIntelCoreUltra2025}
\bibfield{author}{\bibinfo{person}{{Intel Corporation}}.}
  \bibinfo{year}{2025}\natexlab{}.
\newblock \bibinfo{title}{Intel {{Core Ultra Processors}}}.
\newblock
  \bibinfo{howpublished}{\url{https://www.intel.com/content/www/us/en/products/details/processors/core-ultra.html}}.
\newblock


\bibitem[Jacobs et~al\mbox{.}(1991)]%
        {jacobs_adaptive_1991}
\bibfield{author}{\bibinfo{person}{Robert~A. Jacobs},
  \bibinfo{person}{Michael~I. Jordan}, \bibinfo{person}{Steven~J. Nowlan},
  {and} \bibinfo{person}{Geoffrey~E. Hinton}.} \bibinfo{year}{1991}\natexlab{}.
\newblock \showarticletitle{Adaptive Mixtures of Local Experts}.
\newblock  \bibinfo{volume}{3}, \bibinfo{number}{1} (\bibinfo{year}{1991}),
  \bibinfo{pages}{79--87}.
\newblock
\showISSN{0899-7667}
\href{https://doi.org/10.1162/neco.1991.3.1.79}{doi:\nolinkurl{10.1162/neco.1991.3.1.79}}
\newblock
\shownote{Conference Name: Neural Computation}.


\bibitem[Jiang et~al\mbox{.}(2024)]%
        {mixtral}
\bibfield{author}{\bibinfo{person}{Albert~Q Jiang}, \bibinfo{person}{Alexandre
  Sablayrolles}, \bibinfo{person}{Antoine Roux}, \bibinfo{person}{Arthur
  Mensch}, \bibinfo{person}{Blanche Savary}, \bibinfo{person}{Chris Bamford},
  \bibinfo{person}{Devendra~Singh Chaplot}, \bibinfo{person}{Diego de~las
  Casas}, \bibinfo{person}{Emma~Bou Hanna}, \bibinfo{person}{Florian Bressand},
  {et~al\mbox{.}}} \bibinfo{year}{2024}\natexlab{}.
\newblock \showarticletitle{Mixtral of experts}.
\newblock \bibinfo{journal}{\emph{arXiv preprint arXiv:2401.04088}}
  (\bibinfo{year}{2024}).
\newblock


\bibitem[Jimenez et~al\mbox{.}(2024)]%
        {jimenez2024swebench}
\bibfield{author}{\bibinfo{person}{Carlos~E Jimenez}, \bibinfo{person}{John
  Yang}, \bibinfo{person}{Alexander Wettig}, \bibinfo{person}{Shunyu Yao},
  \bibinfo{person}{Kexin Pei}, \bibinfo{person}{Ofir Press}, {and}
  \bibinfo{person}{Karthik~R Narasimhan}.} \bibinfo{year}{2024}\natexlab{}.
\newblock \showarticletitle{{SWE}-bench: Can Language Models Resolve Real-world
  Github Issues?}. In \bibinfo{booktitle}{\emph{The Twelfth International
  Conference on Learning Representations}}.
\newblock
\urldef\tempurl%
\url{https://openreview.net/forum?id=VTF8yNQM66}
\showURL{%
\tempurl}


\bibitem[Kaleem et~al\mbox{.}(2014)]%
        {kaleem2014adaptive}
\bibfield{author}{\bibinfo{person}{Rashid Kaleem}, \bibinfo{person}{Rajkishore
  Barik}, \bibinfo{person}{Tatiana Shpeisman}, \bibinfo{person}{Brian~T Lewis},
  \bibinfo{person}{Chunling Hu}, {and} \bibinfo{person}{Keshav Pingali}.}
  \bibinfo{year}{2014}\natexlab{}.
\newblock \showarticletitle{Adaptive heterogeneous scheduling for integrated
  GPUs}. In \bibinfo{booktitle}{\emph{Proceedings of the 23rd international
  conference on Parallel architectures and compilation}}.
  \bibinfo{pages}{151--162}.
\newblock


\bibitem[Kamahori et~al\mbox{.}(2024)]%
        {kamahori2024fiddler}
\bibfield{author}{\bibinfo{person}{Keisuke Kamahori}, \bibinfo{person}{Yile
  Gu}, \bibinfo{person}{Kan Zhu}, {and} \bibinfo{person}{Baris Kasikci}.}
  \bibinfo{year}{2024}\natexlab{}.
\newblock \bibinfo{title}{Fiddler: CPU-GPU Orchestration for Fast Inference of
  Mixture-of-Experts Models}.
\newblock
\showeprint[arxiv]{2402.07033}~[cs.LG]


\bibitem[Kong et~al\mbox{.}(2024)]%
        {kong2024swapmoe}
\bibfield{author}{\bibinfo{person}{Rui Kong}, \bibinfo{person}{Yuanchun Li},
  \bibinfo{person}{Qingtian Feng}, \bibinfo{person}{Weijun Wang},
  \bibinfo{person}{Xiaozhou Ye}, \bibinfo{person}{Ye Ouyang},
  \bibinfo{person}{Linghe Kong}, {and} \bibinfo{person}{Yunxin Liu}.}
  \bibinfo{year}{2024}\natexlab{}.
\newblock \showarticletitle{SwapMoE: Serving Off-the-shelf MoE-based Large
  Language Models with Tunable Memory Budget}. In
  \bibinfo{booktitle}{\emph{Proceedings of the 62nd Annual Meeting of the
  Association for Computational Linguistics (Volume 1: Long Papers)}}.
  \bibinfo{pages}{6710--6720}.
\newblock


\bibitem[Kwon et~al\mbox{.}(2023)]%
        {vllm}
\bibfield{author}{\bibinfo{person}{Woosuk Kwon}, \bibinfo{person}{Zhuohan Li},
  \bibinfo{person}{Siyuan Zhuang}, \bibinfo{person}{Ying Sheng},
  \bibinfo{person}{Lianmin Zheng}, \bibinfo{person}{Cody~Hao Yu},
  \bibinfo{person}{Joseph Gonzalez}, \bibinfo{person}{Hao Zhang}, {and}
  \bibinfo{person}{Ion Stoica}.} \bibinfo{year}{2023}\natexlab{}.
\newblock \showarticletitle{Efficient Memory Management for Large Language
  Model Serving with {PagedAttention}}. In
  \bibinfo{booktitle}{\emph{Proceedings of the 29th Symposium on Operating
  Systems Principles}} (New York, {NY}, {USA}, 2023-10-23)
  \emph{(\bibinfo{series}{{SOSP} '23})}. \bibinfo{publisher}{Association for
  Computing Machinery}, \bibinfo{pages}{611--626}.
\newblock
\showISBNx{9798400702297}
\href{https://doi.org/10.1145/3600006.3613165}{doi:\nolinkurl{10.1145/3600006.3613165}}


\bibitem[Liu et~al\mbox{.}(2026)]%
        {liu2026multi}
\bibfield{author}{\bibinfo{person}{Mengyang Liu}, \bibinfo{person}{Taozhi
  Chen}, \bibinfo{person}{Zhenhua Xu}, \bibinfo{person}{Xue Jiang}, {and}
  \bibinfo{person}{Yihong Dong}.} \bibinfo{year}{2026}\natexlab{}.
\newblock \showarticletitle{Multi-agent Collaboration with State Management}.
\newblock \bibinfo{journal}{\emph{arXiv preprint arXiv:2605.20563}}
  (\bibinfo{year}{2026}).
\newblock


\bibitem[Markatos and LeBlanc(1994)]%
        {markatos1994using}
\bibfield{author}{\bibinfo{person}{Evangelos~P. Markatos} {and}
  \bibinfo{person}{Thomas~J. LeBlanc}.} \bibinfo{year}{1994}\natexlab{}.
\newblock \showarticletitle{Using processor affinity in loop scheduling on
  shared-memory multiprocessors}.
\newblock \bibinfo{journal}{\emph{IEEE Transactions on Parallel and Distributed
  systems}} \bibinfo{volume}{5}, \bibinfo{number}{4} (\bibinfo{year}{1994}),
  \bibinfo{pages}{379--400}.
\newblock


\bibitem[Polychronopoulos and Kuck(1987)]%
        {polychronopoulos1987guided}
\bibfield{author}{\bibinfo{person}{Constantine~D Polychronopoulos} {and}
  \bibinfo{person}{David~J Kuck}.} \bibinfo{year}{1987}\natexlab{}.
\newblock \showarticletitle{Guided self-scheduling: A practical scheduling
  scheme for parallel supercomputers}.
\newblock \bibinfo{journal}{\emph{IEEE Trans. Comput.}} \bibinfo{volume}{100},
  \bibinfo{number}{12} (\bibinfo{year}{1987}), \bibinfo{pages}{1425--1439}.
\newblock


\bibitem[Power et~al\mbox{.}(2013)]%
        {power2013heterogeneous}
\bibfield{author}{\bibinfo{person}{Jason Power}, \bibinfo{person}{Arkaprava
  Basu}, \bibinfo{person}{Junli Gu}, \bibinfo{person}{Sooraj Puthoor},
  \bibinfo{person}{Bradford~M Beckmann}, \bibinfo{person}{Mark~D Hill},
  \bibinfo{person}{Steven~K Reinhardt}, {and} \bibinfo{person}{David~A Wood}.}
  \bibinfo{year}{2013}\natexlab{}.
\newblock \showarticletitle{Heterogeneous system coherence for integrated
  CPU-GPU systems}. In \bibinfo{booktitle}{\emph{Proceedings of the 46th Annual
  IEEE/ACM International Symposium on Microarchitecture}}.
  \bibinfo{pages}{457--467}.
\newblock


\bibitem[Prokopec and Odersky(2013)]%
        {prokopec2013near}
\bibfield{author}{\bibinfo{person}{Aleksandar Prokopec} {and}
  \bibinfo{person}{Martin Odersky}.} \bibinfo{year}{2013}\natexlab{}.
\newblock \showarticletitle{Near optimal work-stealing tree scheduler for
  highly irregular data-parallel workloads}. In
  \bibinfo{booktitle}{\emph{International Workshop on Languages and Compilers
  for Parallel Computing}}. Springer, \bibinfo{pages}{55--86}.
\newblock


\bibitem[Qu et~al\mbox{.}(2025)]%
        {qu2025mobile}
\bibfield{author}{\bibinfo{person}{Guanqiao Qu}, \bibinfo{person}{Qiyuan Chen},
  \bibinfo{person}{Wei Wei}, \bibinfo{person}{Zheng Lin},
  \bibinfo{person}{Xianhao Chen}, {and} \bibinfo{person}{Kaibin Huang}.}
  \bibinfo{year}{2025}\natexlab{}.
\newblock \showarticletitle{Mobile edge intelligence for large language models:
  A contemporary survey}.
\newblock \bibinfo{journal}{\emph{IEEE Communications Surveys \& Tutorials}}
  \bibinfo{volume}{27}, \bibinfo{number}{6} (\bibinfo{year}{2025}),
  \bibinfo{pages}{3820--3860}.
\newblock


\bibitem[{Qualcomm}(2025)]%
        {qualcommLaptopsPoweredSnapdragon}
\bibfield{author}{\bibinfo{person}{{Qualcomm}}.}
  \bibinfo{year}{2025}\natexlab{}.
\newblock \bibinfo{title}{Laptops Powered by {{Snapdragon}}}.
\newblock
  \bibinfo{howpublished}{\url{https://www.qualcomm.com/snapdragon/laptops-and-tablets}}.
\newblock


\bibitem[Skende et~al\mbox{.}(2026)]%
        {skende171NVIDIAGB102026}
\bibfield{author}{\bibinfo{person}{A. Skende}, \bibinfo{person}{G. Rosseel},
  {and} \bibinfo{person}{N. Pinckney}.} \bibinfo{year}{2026}\natexlab{}.
\newblock \showarticletitle{17.1 NVIDIA GB10: SoC Built for AI Acceleration}.
  In \bibinfo{booktitle}{\emph{2026 IEEE International Solid-State Circuits
  Conference (ISSCC)}}, Vol.~\bibinfo{volume}{69}. \bibinfo{pages}{298--300}.
\newblock
\href{https://doi.org/10.1109/ISSCC49663.2026.11409351}{doi:\nolinkurl{10.1109/ISSCC49663.2026.11409351}}


\bibitem[Song et~al\mbox{.}(2024)]%
        {song2024promoe}
\bibfield{author}{\bibinfo{person}{Xiaoniu Song}, \bibinfo{person}{Zihang
  Zhong}, \bibinfo{person}{Rong Chen}, {and} \bibinfo{person}{Haibo Chen}.}
  \bibinfo{year}{2024}\natexlab{}.
\newblock \showarticletitle{Promoe: Fast moe-based llm serving using proactive
  caching}.
\newblock \bibinfo{journal}{\emph{arXiv preprint arXiv:2410.22134}}
  (\bibinfo{year}{2024}).
\newblock


\bibitem[Song et~al\mbox{.}(2023)]%
        {song_powerinfer_2023}
\bibfield{author}{\bibinfo{person}{Yixin Song}, \bibinfo{person}{Zeyu Mi},
  \bibinfo{person}{Haotong Xie}, {and} \bibinfo{person}{Haibo Chen}.}
  \bibinfo{year}{2023}\natexlab{}.
\newblock \bibinfo{booktitle}{\emph{{PowerInfer}: Fast Large Language Model
  Serving with a Consumer-grade {GPU}}}.
\newblock
\urldef\tempurl%
\url{https://arxiv.org/abs/2312.12456v1}
\showURL{%
\tempurl}


\bibitem[Tang et~al\mbox{.}(2026)]%
        {MoE-APEX}
\bibfield{author}{\bibinfo{person}{Peng Tang}, \bibinfo{person}{Jiacheng Liu},
  \bibinfo{person}{Xiaofeng Hou}, \bibinfo{person}{Yifei Pu},
  \bibinfo{person}{Jing Wang}, \bibinfo{person}{Pheng-Ann Heng},
  \bibinfo{person}{Chao Li}, {and} \bibinfo{person}{Minyi Guo}.}
  \bibinfo{year}{2026}\natexlab{}.
\newblock \showarticletitle{MoE-APEX: An Efficient MoE Inference System with
  Adaptive Precision Expert Offloading}. In
  \bibinfo{booktitle}{\emph{Proceedings of the 31st ACM International
  Conference on Architectural Support for Programming Languages and Operating
  Systems, Volume 2}} (USA) \emph{(\bibinfo{series}{ASPLOS '26})}.
  \bibinfo{publisher}{Association for Computing Machinery},
  \bibinfo{address}{New York, NY, USA}, \bibinfo{pages}{1185–1200}.
\newblock
\showISBNx{9798400723599}
\href{https://doi.org/10.1145/3779212.3790187}{doi:\nolinkurl{10.1145/3779212.3790187}}


\bibitem[Team(2025)]%
        {qwen3technicalreport}
\bibfield{author}{\bibinfo{person}{Qwen Team}.}
  \bibinfo{year}{2025}\natexlab{}.
\newblock \bibinfo{title}{Qwen3 Technical Report}.
\newblock
\showeprint[arxiv]{2505.09388}~[cs.CL]
\urldef\tempurl%
\url{https://arxiv.org/abs/2505.09388}
\showURL{%
\tempurl}


\bibitem[Vandierendonck et~al\mbox{.}(2011)]%
        {vandierendonck2011unified}
\bibfield{author}{\bibinfo{person}{Hans Vandierendonck},
  \bibinfo{person}{George Tzenakis}, {and} \bibinfo{person}{Dimitrios~S
  Nikolopoulos}.} \bibinfo{year}{2011}\natexlab{}.
\newblock \showarticletitle{A unified scheduler for recursive and task dataflow
  parallelism}. In \bibinfo{booktitle}{\emph{2011 International Conference on
  Parallel Architectures and Compilation Techniques}}. IEEE,
  \bibinfo{pages}{1--11}.
\newblock


\bibitem[Vet et~al\mbox{.}(2012)]%
        {vet2012multigrain}
\bibfield{author}{\bibinfo{person}{Jean-Yves Vet}, \bibinfo{person}{Patrick
  Carribault}, {and} \bibinfo{person}{Albert Cohen}.}
  \bibinfo{year}{2012}\natexlab{}.
\newblock \showarticletitle{Multigrain affinity for heterogeneous work
  stealing}. In \bibinfo{booktitle}{\emph{Programmability Issues for
  Heterogeneous Multicores}}.
\newblock


\bibitem[Wei et~al\mbox{.}(2025)]%
        {wei2025ghidorah}
\bibfield{author}{\bibinfo{person}{Jinhui Wei}, \bibinfo{person}{Ye Huang},
  \bibinfo{person}{Yuhui Zhou}, \bibinfo{person}{Jiazhi Jiang}, {and}
  \bibinfo{person}{Jiangsu Du}.} \bibinfo{year}{2025}\natexlab{}.
\newblock \showarticletitle{Ghidorah: Fast LLM inference on edge with
  speculative decoding and hetero-core parallelism}. In
  \bibinfo{booktitle}{\emph{2025 IEEE 43rd International Conference on Computer
  Design (ICCD)}}. IEEE, \bibinfo{pages}{253--260}.
\newblock


\bibitem[Xu et~al\mbox{.}(2024)]%
        {xu2024device}
\bibfield{author}{\bibinfo{person}{Jiajun Xu}, \bibinfo{person}{Zhiyuan Li},
  \bibinfo{person}{Wei Chen}, \bibinfo{person}{Qun Wang}, \bibinfo{person}{Xin
  Gao}, \bibinfo{person}{Qi Cai}, {and} \bibinfo{person}{Ziyuan Ling}.}
  \bibinfo{year}{2024}\natexlab{}.
\newblock \showarticletitle{On-device language models: A comprehensive review}.
\newblock \bibinfo{journal}{\emph{arXiv preprint arXiv:2409.00088}}
  (\bibinfo{year}{2024}).
\newblock


\bibitem[Xue et~al\mbox{.}(2024)]%
        {moe-infinity}
\bibfield{author}{\bibinfo{person}{Leyang Xue}, \bibinfo{person}{Yao Fu},
  \bibinfo{person}{Zhan Lu}, \bibinfo{person}{Luo Mai}, {and}
  \bibinfo{person}{Mahesh Marina}.} \bibinfo{year}{2024}\natexlab{}.
\newblock \bibinfo{title}{{MoE}-Infinity: Offloading-Efficient {MoE} Model
  Serving}.
\newblock
\showeprint[arxiv]{2401.14361}
\href{https://doi.org/10.48550/arXiv.2401.14361}{doi:\nolinkurl{10.48550/arXiv.2401.14361}}


\bibitem[Yang et~al\mbox{.}(2024b)]%
        {yang_qwen2_2024}
\bibfield{author}{\bibinfo{person}{An Yang}, \bibinfo{person}{Baosong Yang},
  \bibinfo{person}{Binyuan Hui}, \bibinfo{person}{Bo Zheng},
  \bibinfo{person}{Bowen Yu}, \bibinfo{person}{Chang Zhou},
  \bibinfo{person}{Chengpeng Li}, \bibinfo{person}{Chengyuan Li},
  \bibinfo{person}{Dayiheng Liu}, \bibinfo{person}{Fei Huang},
  \bibinfo{person}{Guanting Dong}, \bibinfo{person}{Haoran Wei},
  \bibinfo{person}{Huan Lin}, \bibinfo{person}{Jialong Tang},
  \bibinfo{person}{Jialin Wang}, \bibinfo{person}{Jian Yang},
  \bibinfo{person}{Jianhong Tu}, \bibinfo{person}{Jianwei Zhang},
  \bibinfo{person}{Jianxin Ma}, \bibinfo{person}{Jianxin Yang},
  \bibinfo{person}{Jin Xu}, \bibinfo{person}{Jingren Zhou},
  \bibinfo{person}{Jinze Bai}, \bibinfo{person}{Jinzheng He},
  \bibinfo{person}{Junyang Lin}, \bibinfo{person}{Kai Dang},
  \bibinfo{person}{Keming Lu}, \bibinfo{person}{Keqin Chen},
  \bibinfo{person}{Kexin Yang}, \bibinfo{person}{Mei Li},
  \bibinfo{person}{Mingfeng Xue}, \bibinfo{person}{Na Ni}, \bibinfo{person}{Pei
  Zhang}, \bibinfo{person}{Peng Wang}, \bibinfo{person}{Ru Peng},
  \bibinfo{person}{Rui Men}, \bibinfo{person}{Ruize Gao},
  \bibinfo{person}{Runji Lin}, \bibinfo{person}{Shijie Wang},
  \bibinfo{person}{Shuai Bai}, \bibinfo{person}{Sinan Tan},
  \bibinfo{person}{Tianhang Zhu}, \bibinfo{person}{Tianhao Li},
  \bibinfo{person}{Tianyu Liu}, \bibinfo{person}{Wenbin Ge},
  \bibinfo{person}{Xiaodong Deng}, \bibinfo{person}{Xiaohuan Zhou},
  \bibinfo{person}{Xingzhang Ren}, \bibinfo{person}{Xinyu Zhang},
  \bibinfo{person}{Xipin Wei}, \bibinfo{person}{Xuancheng Ren},
  \bibinfo{person}{Xuejing Liu}, \bibinfo{person}{Yang Fan},
  \bibinfo{person}{Yang Yao}, \bibinfo{person}{Yichang Zhang},
  \bibinfo{person}{Yu Wan}, \bibinfo{person}{Yunfei Chu},
  \bibinfo{person}{Yuqiong Liu}, \bibinfo{person}{Zeyu Cui},
  \bibinfo{person}{Zhenru Zhang}, \bibinfo{person}{Zhifang Guo}, {and}
  \bibinfo{person}{Zhihao Fan}.} \bibinfo{year}{2024}\natexlab{b}.
\newblock \bibinfo{title}{Qwen2 Technical Report}.
\newblock
\showeprint[arxiv]{2407.10671}
\href{https://doi.org/10.48550/arXiv.2407.10671}{doi:\nolinkurl{10.48550/arXiv.2407.10671}}
\newblock
\shownote{version: 4}.


\bibitem[Yang et~al\mbox{.}(2024a)]%
        {yang2024sweagentagentcomputerinterfacesenable}
\bibfield{author}{\bibinfo{person}{John Yang}, \bibinfo{person}{Carlos~E.
  Jimenez}, \bibinfo{person}{Alexander Wettig}, \bibinfo{person}{Kilian
  Lieret}, \bibinfo{person}{Shunyu Yao}, \bibinfo{person}{Karthik Narasimhan},
  {and} \bibinfo{person}{Ofir Press}.} \bibinfo{year}{2024}\natexlab{a}.
\newblock \bibinfo{title}{SWE-agent: Agent-Computer Interfaces Enable Automated
  Software Engineering}.
\newblock
\showeprint[arxiv]{2405.15793}~[cs.SE]
\urldef\tempurl%
\url{https://arxiv.org/abs/2405.15793}
\showURL{%
\tempurl}


\bibitem[Yang et~al\mbox{.}(2027)]%
        {yang2027libumsh}
\bibfield{author}{\bibinfo{person}{Jinrong Yang}, \bibinfo{person}{Zimeng
  Wang}, \bibinfo{person}{Rong Chen}, {and} \bibinfo{person}{Haibo Chen}.}
  \bibinfo{year}{2027}\natexlab{}.
\newblock \showarticletitle{Characterizing CPU-GPU Memory Sharing over Unified
  Memory Architecture}. In \bibinfo{booktitle}{\emph{Proceedings of the 22nd
  Conference on Computer Systems}} \emph{(\bibinfo{series}{EuroSys '27})}.
\newblock


\bibitem[Yang et~al\mbox{.}(2026)]%
        {yang2026freetoken}
\bibfield{author}{\bibinfo{person}{Shuo Yang}, \bibinfo{person}{Xiaoze Fan},
  \bibinfo{person}{Melissa Pan}, \bibinfo{person}{Haocheng Xi},
  \bibinfo{person}{Zhe Wang}, \bibinfo{person}{Shanlin Sun},
  \bibinfo{person}{Kurt Keutzer}, \bibinfo{person}{Song Han},
  \bibinfo{person}{Matei Zaharia}, \bibinfo{person}{Chenfeng Xu},
  {et~al\mbox{.}}} \bibinfo{year}{2026}\natexlab{}.
\newblock \showarticletitle{FreeToken: Efficient Edge-Native MoE Serving with
  Bandwidth-Adaptive Execution}.
\newblock \bibinfo{journal}{\emph{arXiv preprint arXiv:2608.16157}}
  (\bibinfo{year}{2026}).
\newblock


\bibitem[Yi et~al\mbox{.}(2023)]%
        {edgemoe}
\bibfield{author}{\bibinfo{person}{Rongjie Yi}, \bibinfo{person}{Liwei Guo},
  \bibinfo{person}{Shiyun Wei}, \bibinfo{person}{Ao Zhou},
  \bibinfo{person}{Shangguang Wang}, {and} \bibinfo{person}{Mengwei Xu}.}
  \bibinfo{year}{2023}\natexlab{}.
\newblock \bibinfo{title}{{EdgeMoE}: Fast On-Device Inference of {MoE}-based
  Large Language Models}.
\newblock
\showeprint[arxiv]{2308.14352}
\href{https://doi.org/10.48550/arXiv.2308.14352}{doi:\nolinkurl{10.48550/arXiv.2308.14352}}


\bibitem[Yu et~al\mbox{.}(2025)]%
        {LayerScope}
\bibfield{author}{\bibinfo{person}{Enda Yu}, \bibinfo{person}{Dezun Dong},
  \bibinfo{person}{Zhaoning Zhang}, \bibinfo{person}{Zhe Bai},
  \bibinfo{person}{Weiling Yang}, \bibinfo{person}{Haojie Wang},
  \bibinfo{person}{Dongsheng Li}, \bibinfo{person}{Yong Wu}, {and}
  \bibinfo{person}{Xiangke Liao}.} \bibinfo{year}{2025}\natexlab{}.
\newblock \showarticletitle{LayerScope: Predictive Cross-Layer Scheduling for
  Efficient Multi-Batch MoE Inference on Legacy Servers}.
\newblock
\urldef\tempurl%
\url{https://api.semanticscholar.org/CorpusID:281676157}
\showURL{%
\tempurl}


\bibitem[Yu et~al\mbox{.}(2022)]%
        {orca}
\bibfield{author}{\bibinfo{person}{Gyeong-In Yu}, \bibinfo{person}{Joo~Seong
  Jeong}, \bibinfo{person}{Geon-Woo Kim}, \bibinfo{person}{Soojeong Kim}, {and}
  \bibinfo{person}{Byung-Gon Chun}.} \bibinfo{year}{2022}\natexlab{}.
\newblock \showarticletitle{Orca: A Distributed Serving System for
  {Transformer-Based} Generative Models}. In \bibinfo{booktitle}{\emph{16th
  USENIX Symposium on Operating Systems Design and Implementation (OSDI 22)}}.
  \bibinfo{publisher}{USENIX Association}, \bibinfo{address}{Carlsbad, CA},
  \bibinfo{pages}{521--538}.
\newblock
\showISBNx{978-1-939133-28-1}
\urldef\tempurl%
\url{https://www.usenix.org/conference/osdi22/presentation/yu}
\showURL{%
\tempurl}


\bibitem[Yu et~al\mbox{.}(2026)]%
        {yu2026taming}
\bibfield{author}{\bibinfo{person}{Hanfei Yu}, \bibinfo{person}{Xingqi Cui},
  \bibinfo{person}{Hong Zhang}, \bibinfo{person}{Hao Wang}, {and}
  \bibinfo{person}{Hao Wang}.} \bibinfo{year}{2026}\natexlab{}.
\newblock \showarticletitle{Taming Latency-Memory Trade-Off in MoE-Based LLM
  Serving via Fine-Grained Expert Offloading}. In
  \bibinfo{booktitle}{\emph{Proceedings of the 21st European Conference on
  Computer Systems}}. \bibinfo{pages}{176--191}.
\newblock


\bibitem[Zaharia et~al\mbox{.}(2008)]%
        {zaharia2008improving}
\bibfield{author}{\bibinfo{person}{Matei Zaharia}, \bibinfo{person}{Andy
  Konwinski}, \bibinfo{person}{Anthony~D Joseph}, \bibinfo{person}{Randy~H
  Katz}, {and} \bibinfo{person}{Ion Stoica}.} \bibinfo{year}{2008}\natexlab{}.
\newblock \showarticletitle{Improving MapReduce performance in heterogeneous
  environments.}. In \bibinfo{booktitle}{\emph{Osdi}},
  Vol.~\bibinfo{volume}{8}. \bibinfo{pages}{7}.
\newblock


\bibitem[Zhang et~al\mbox{.}(2025)]%
        {zhang2025optimizing}
\bibfield{author}{\bibinfo{person}{Enhao Zhang}, \bibinfo{person}{Erkang Zhu},
  \bibinfo{person}{Gagan Bansal}, \bibinfo{person}{Adam Fourney},
  \bibinfo{person}{Hussein Mozannar}, {and} \bibinfo{person}{Jack Gerrits}.}
  \bibinfo{year}{2025}\natexlab{}.
\newblock \showarticletitle{Optimizing sequential multi-step tasks with
  parallel llm agents}.
\newblock \bibinfo{journal}{\emph{arXiv preprint arXiv:2507.08944}}
  (\bibinfo{year}{2025}).
\newblock


\bibitem[Zhang et~al\mbox{.}(2024)]%
        {zhang2024autocoderoverautonomousprogramimprovement}
\bibfield{author}{\bibinfo{person}{Yuntong Zhang}, \bibinfo{person}{Haifeng
  Ruan}, \bibinfo{person}{Zhiyu Fan}, {and} \bibinfo{person}{Abhik
  Roychoudhury}.} \bibinfo{year}{2024}\natexlab{}.
\newblock \bibinfo{title}{AutoCodeRover: Autonomous Program Improvement}.
\newblock
\showeprint[arxiv]{2404.05427}~[cs.SE]
\urldef\tempurl%
\url{https://arxiv.org/abs/2404.05427}
\showURL{%
\tempurl}


\bibitem[Zhong et~al\mbox{.}(2025)]%
        {hybrimoe}
\bibfield{author}{\bibinfo{person}{Shuzhang Zhong}, \bibinfo{person}{Yanfan
  Sun}, \bibinfo{person}{Ling Liang}, \bibinfo{person}{Runsheng Wang},
  \bibinfo{person}{Ru Huang}, {and} \bibinfo{person}{Meng Li}.}
  \bibinfo{year}{2025}\natexlab{}.
\newblock \showarticletitle{Hybrimoe: Hybrid cpu-gpu scheduling and cache
  management for efficient moe inference}. In \bibinfo{booktitle}{\emph{2025
  62nd ACM/IEEE Design Automation Conference (DAC)}}. IEEE,
  \bibinfo{pages}{1--7}.
\newblock


\bibitem[Zhong et~al\mbox{.}(2024)]%
        {zhong2024distserve}
\bibfield{author}{\bibinfo{person}{Yinmin Zhong}, \bibinfo{person}{Shengyu
  Liu}, \bibinfo{person}{Junda Chen}, \bibinfo{person}{Jianbo Hu},
  \bibinfo{person}{Yibo Zhu}, \bibinfo{person}{Xuanzhe Liu},
  \bibinfo{person}{Xin Jin}, {and} \bibinfo{person}{Hao Zhang}.}
  \bibinfo{year}{2024}\natexlab{}.
\newblock \showarticletitle{$\{$DistServe$\}$: Disaggregating prefill and
  decoding for goodput-optimized large language model serving}. In
  \bibinfo{booktitle}{\emph{18th USENIX symposium on operating systems design
  and implementation (OSDI 24)}}. \bibinfo{pages}{193--210}.
\newblock


\bibitem[Zhu et~al\mbox{.}(2026)]%
        {zhu2026smoealgorithmsystemcodesignpushing}
\bibfield{author}{\bibinfo{person}{Guoying Zhu}, \bibinfo{person}{Meng Li},
  \bibinfo{person}{Haipeng Dai}, \bibinfo{person}{Xuechen Liu},
  \bibinfo{person}{Weijun Wang}, \bibinfo{person}{Keran Li},
  \bibinfo{person}{Jun xiao}, \bibinfo{person}{Ligeng Chen}, {and}
  \bibinfo{person}{Wei Wang}.} \bibinfo{year}{2026}\natexlab{}.
\newblock \bibinfo{title}{SMoE: An Algorithm-System Co-Design for Pushing MoE
  to the Edge via Expert Substitution}.
\newblock
\showeprint[arxiv]{2508.18983}~[cs.AI]
\urldef\tempurl%
\url{https://arxiv.org/abs/2508.18983}
\showURL{%
\tempurl}


\end{thebibliography}
}

\end{document}